\documentclass[%
 reprint,
 amsmath,amssymb,
 aps,
]{revtex4-2}

\usepackage{graphicx}
\usepackage{dcolumn}
\usepackage{bm}
\usepackage{xcolor}
\usepackage{mathrsfs}
\usepackage[normalem]{ulem}

\begin{document}


\title{Transport of condensing droplets in Taylor-Green vortex flow in the presence of thermal noise}

\author{Anu V. S. Nath}
\email{am18d701@smail.iitm.ac.in}
\author{Anubhab Roy}%
 \email{anubhab@iitm.ac.in}
\affiliation{Department of Applied Mechanics,
Indian Institute of Technology Madras, Chennai 600036}

\author{Rama Govindarajan}
\email{rama@icts.res.in}
\affiliation{International Centre for Theoretical Sciences,
Tata Institute of Fundamental Research, Bengaluru 560089
}

\author{S. Ravichandran}
\email{ravichandran@su.se}
\affiliation{ Nordita, KTH Royal Institute of Technology and Stockholm University,
Stockholm, Sweden, SE-10691}


\begin{abstract}
We study the role of phase change and thermal noise in particle transport in turbulent flows. We employ a toy model to extract the main physics: condensing droplets are modelled as heavy particles which grow in size, the ambient flow is modelled as a two-dimensional Taylor-Green (TG) flow consisting of an array of vortices delineated by separatrices, and thermal noise are modelled as uncorrelated Gaussian white noise. In general, heavy inertial particles are centrifuged out of regions of high vorticity and into regions of high strain. In cellular flows, we find, in agreement with earlier results, that droplets with Stokes numbers smaller than a critical value, $St < St_{\rm{cr}}$, remain trapped in the vortices in which they are initialised, while larger droplets move ballistically away from their initial positions by crossing separatrices. We independently vary the P\'eclet number $Pe$ characterising the amplitude of thermal noise and the condensation rate $\Pi$ to study their effects on the critical Stokes number for droplet trapping, as well as on the final states of motion of the droplets. We find that the imposition of thermal noise, or of a finite condensation rate, allows droplets of $St < St_{\rm{cr}}$ to leave their initial vortices. We find that the effects of thermal noise become negligible for growing droplets, and that growing droplets achieve ballistic motion when their Stokes numbers become $\mathcal{O}(1)$. We also find an intermediate regime prior to attaining the ballistic state, in which droplets move diffusively away from their initial vortices in the presence of thermal noise.

\end{abstract}

\maketitle

\section{Introduction}

Fluid flows in which solid particles, liquid droplets or gas bubbles of a different material are suspended are the rule rather than the exception in natural and industrial settings \cite{clift_grace}. The suspended entities could range from a few micrometres in size (water droplets in clouds) to several kilometres (asteroids in the interstellar medium). Such suspended `particles' are advected by the flow, but, due to their finite size, do not necessarily follow fluid streamlines. As a result, the dynamics of the suspended (`inertial') particles could be qualitatively different from that of the carrier fluid. For instance, particles much denser than the fluid (`heavy' inertial particles) are centrifuged out of vortical regions and cluster in strain-dominated regions in the flow \cite{maxey1987motion}. Turbulent flows are, in general, chaotic tangles of vortex tubes and sheets \cite{vincent_meneguzzi_1994}, and heavy inertial particles suspended in turbulent flow are known to cluster onto fractal attractors \cite{bec2003fractal}. The clustering of heavy inertial particles in turbulent flow has been studied theoretically, experimentally, and numerically \cite{chen2006turbulent,angilella2010dust,monchaux2012analyzing,petersen2019experimental,yoshimoto2007self,bragg2014new,goto2006self,saw2012spatial,baker2017coherent,ravichandran2014attracting}, and has been reviewed in, e.g., ref. \cite{gustavsson2016statistical}.

The dynamics of individual inertial particles can give rise to multivalued particle velocities at a given spatial location and time, even when they are suspended in an incompressible fluid. These events give rise to folds in particle-velocity space, commonly referred to as `caustics' \cite{wilkinson2005caustics,gustavsson2012inertial,bec2010intermittency,gustavsson2014tumbling,ravichandran2015caustics}. Caustics lead to enhanced clustering, and are known to lead to higher collision rates between inertial particles \cite{gustavsson2016statistical1,deepu2017caustics}.

Numerical studies of turbulent suspensions model the suspended phase either as a continuum (the Eulerian approach) or as discrete units which need to be tracked individually (the Lagrangian approach) \cite{elghobashi1994predicting}. These models can then be coupled to a suitable solver for the Navier-Stokes equations governing the dynamics of the fluid. 

When the volume or mass fraction of the suspended phase is sufficiently small, the feedback from the particles on the flow can be neglected. This allows the dynamics of such particles to be studied separately from the flow, for example, by using publicly available data sets of turbulent flow (like the Johns Hopkins Turbulence Database, JHTDB. See, e.g. \cite{candelier2016angular} who study the orientation dynamics of asymmetric particles). However, when the volume fractions are not negligible, the feedback from the particles on the flow cannot be neglected, and the dynamics of the flow and the suspended particles have to be studied simultaneously \cite{ richter2014modification, muramulla2020disruption}. 

When the suspended phase is made of liquid droplets rather than solid particles, these droplets can also qualitatively change the flow through the exchange of mass or energy. For example, the evaporation of water droplets formed by wave-breaking at the ocean surface generates cloud condensation nuclei \cite{veron2015ocean}, while the latent heat release accompanying the growth of water droplets drives the dynamics in clouds \cite{narasimha2012cumulus}. The fluid dynamics of respiratory events, relevant to the ongoing global pandemic, also involve evaporating droplets suspended in turbulent flows (e.g. \cite{lohse2021,rosti2021turbulence,diwan2020,singhal2021}). 

It is reasonable to assume that small suspended droplets are spherical. The mass transfer may be assumed to occur diffusively, leading to an analytical expression for the rate of growth of the droplets (see \cite{pruppacher2010microstructure} for a detailed derivation). Higher order corrections accounting for advective mass transfer have also been proposed and used (e.g. \cite{andreas1989thermal,helgans2016turbulent,veron2015ocean,lohse2021,rosti2021turbulence}). These higher order corrections may be neglected when the droplets are sufficiently small. In addition, if the suspended droplets are small but larger than a critical radius called the Köhler radius, effects due to the curvature and salt concentration on the rate of growth may also be neglected, and the growth rate takes on a simple form (\cite{shaw1998preferential,vaillancourt2002microscopic,sardina2015continuous}; see section \ref{sec2}).

In addition to the systematic forces described above, suspended particles may also experience stochastic forces from collisions with fluid molecules due to thermal noises/fluctuations. For sufficiently small particles, this leads to the well known phenomenon of `Brownian motion', a diffusive motion. For larger particles, the effects of the thermal noise are negligible. Studies of the dynamics of particles in the size range where both systematic inertial effects as well as the effects of stochastic forces are relevant are relatively rare, and include studies in simple shear flows (\cite{drossinos2005brownian}), Taylor-Green vortices (\cite{renaud2020dispersion,pavliotis2005periodic}), and turbulent flows (\cite{elperin1996turbulent}).

Renaud \& Vanneste~\cite{renaud2020dispersion} quantified the thermal diffusion of particles using an effective diffusivity $D_{\text{eff}}$, for heavy and light inertial particles for various ranges of the Stokes number $St$ and the Péclet number $Pe$. The Stokes number, $St=\tau_p/\tau_f$, where $\tau_p$ and $\tau_f$ are particle relaxation time scale and flow time scale respectively, is a measure of the inertia of a particle. The P\'eclet number is given by $Pe=\tau_d/\tau_f$, where $\tau_d = L_f^2/D_E$ is the diffusion time scale, $L_f$ is the length scale of the flow and $D_E$ is the Einstein diffusivity of a particle due to thermal noise (see section \ref{sec2}), and is related to the strength of the thermal noise. Both $St$ and $Pe$ are increasing functions of the particle size. Thus in flows where the suspended droplets grow or shrink due to phase change, $St$ and $Pe$ are functions of time. The effects of thermal diffusion are important in the early evolution of droplets growing by condensation, and become negligible as the droplets become sufficiently large. 

Here, we study the combined effects of growth by condensation and thermal diffusion on water droplets in clouds. We use an array of TG vortices as a `toy model' for the highly turbulent flow in clouds \cite{grabowski2013growth}. This approach is in the same vein as studies using model flows as proxy for turbulent environments (e.g. \cite{popli2020pattern,gustavsson2012inertial} where the turbulence is modelled as a superposition of Fourier modes). 

The dispersion of inertial particles in cellular flows has been studied without \cite{crisanti1990passive,crisanti1992dynamics,wang1992chaotic,jayaram2020clustering,baggaley2016stability,samant2021dynamic} and with gravity \cite{maxey1986gravitational,rubin1995settling,bergougnoux2014motion}. These studies find that, depending on their Stokes number and density ratio, inertial particles can display chaotic dynamics even in non-chaotic flows. In fact, in time-periodic flows, even tracer particles can display chaotic dynamics \cite{rg2002}.

Wang et al.~\cite{wang1992chaotic} found that large$-St$ inertial particles suspended in a TG flow undergo periodic zig-zag motion along open trajectories in the long-time limit. Here we call this kind of motion `ballistic' (see sections \ref{sec4} and \ref{sec5}). 
In contrast, Renaud et al.~\cite{renaud2020dispersion} found that inertial particles in TG flow with thermal noise behave diffusively at long times,  when the initial conditions are forgotten. We examine the competition between these two non-additive effects on droplets. We provide a supersaturated environment in which our droplets can condense, so both $St$ and $Pe$ increases with time.

The remainder of this paper is organised as follows. In Sec.~\ref{sec2} we set down the general formulation used in this study. We then revisit the dynamics of inertial particles in the TG flow in Sec.~\ref{sec3}, and the role of thermal diffusion in inertial particle dynamics in Sec.~\ref{sec4}. We examine the effects of condensation growth of droplets on their dispersion in Sec.~\ref{sec5}. We study the combined effects of condensation growth and thermal diffusion  in Sec. \ref{sec6}. We conclude in Sec.~\ref{sec7}.

\section{\label{sec2} Problem formulation}
The motion of suspended droplets is governed by the exchange of momentum, mass and heat between the droplets and the ambient fluid. Here, we model the momentum transfer using the simple form of the Langevin equation,
\begin{eqnarray}
    \frac{{\rm d} \textbf{v}_p}{{\rm d} t} = \frac{\textbf{u}(\textbf{x}_p)-\textbf{v}_p}{\tau_p} +\frac{\sqrt{2\, D_E}}{\tau_p}\, \bm{\eta}(t)~,
    \label{eq1}
\end{eqnarray}
where $\textbf{v}_p$ is the Lagrangian velocity of the droplet, $\textbf{u}$ is the ambient flow velocity at the droplet location $\textbf{x}_p$ and $\bm{\eta}$ is the stochastic forcing due to thermal noise, whose form will be discussed later. The relaxation time $\tau_p = 2\, r^2\, \rho/(9\, \mu_f)$, is the time scale on which the velocity of the droplet relaxes to the fluid velocity, where  $\rho$ and $r$ are the instantaneous density and radius of the droplet, and $\mu_f$ is the dynamic viscosity of the ambient fluid (air). The Einstein diffusivity ($D_E$) depends on the instantaneous size of droplets, and so its value evolves over time. In Eq.~(\ref{eq1}), it is assumed that the dominant balance in the droplet dynamics is between the acceleration of the droplet and the Stokes drag and stochastic forces on the droplet, and the effects of added mass, the Saffman lift force and the Basset history force are neglected (see \cite{maxey1983equation}). This is justified in the heavy-particle limit, i.e. when the density ratio of droplet to air is large ($\rho/\rho_f \sim \mathcal{O}(10^3) \gg 1$) (see \cite{bergougnoux2014motion}). We have also neglected gravity, hydrodynamic interactions and collisions between the droplets. These may not in general be negligible, but are fair assumptions on a horizontal plane, for particle sizes much smaller than flow length scales, and dilute suspensions respectively. Moreover, our focus is on the effects of condensation and thermal diffusion. We consider small droplets to begin with, which are in Stokes flow relative to the ambient fluid, so we do not have included drag corrections based on Reynolds number. Additionally, we assume that the droplets remain spherical at all times and, therefore, that their angular dynamics need not be considered.

In supersaturated ambients, suspended water droplets grow by the diffusion of water molecules towards their surfaces, and their subsequent adsorption. An expression for the diffusive growth rate of water, accounting for the effects of solutes present in the water droplet, as well as the effects of the finite radius of curvature may be found in \cite{pruppacher2010microstructure}. For droplets that are sufficiently large ($r > 5 \,\mu m$) for these effects to be neglected, the expression for the growth rate takes on a simple form (see, e.g., \cite{yau1996short,shaw1998preferential,vaillancourt2002microscopic,sardina2015continuous}), viz
\begin{equation} 
\frac{{\rm d} r^2}{{\rm d}t} = 2\,\xi~.
\label{eq2}
\end{equation}
The parameter $\xi$ is proportional to the vapour pressure difference between droplet surface and the ambient, which is assumed to be constant here. This is a fair assumption in a dilute suspension, since the ambient temperature and water vapour concentration will not change significantly due to condensation events. Eq.~(\ref{eq2}) may be then integrated to obtain the instantaneous radius as $r(t) = \sqrt{r_0^2+2\, \xi\, t}$, where $r_0=r(t=0)$ is the initial radius of the droplet, and we refer to this as the `parabolic growth model'. The parameter $\xi$ can be written as $\xi_1\, s$, where $s$ is the ambient supersaturation, and $\xi_1$ is proportional to the mass transfer coefficient.
 
Particles suspended in a quiescent ambient which is in thermal equilibrium can nevertheless experience random collisions with molecules of the fluid, leading to stochastic motion of the particle. This was first observed by Robert Brown in 1827 for pollen grains in water. In 1905, Albert Einstein used a molecular approach to derive an expression (called the Einstein-Smoluchowski relation) for the diffusivity (called the Einstein diffusivity or the Brownian diffusivity), $D_{E} = k_B\, T/(6\, \pi\, \mu\, a)$ of such particles, where $k_B$ is the Boltzmann constant, $T$ is the temperature of the system at equilibrium, $\mu$ is the dynamic viscosity of quiescent ambient fluid and $a$ is the radius of the spherical particles. Ornstein \& Uhlenbeck~\cite{uhlenbeck1930theory} showed that by modelling the stochastic thermal noise ($\bm{\eta}$) as a simple Gaussian `white-noise', in the vanishing limit of particle inertia, Einstein diffusivity is recovered. The white-noise is an uncorrelated random signal which has zero mean ($\langle \bm{\eta(t)}\rangle = \textbf{0}$) and an auto-correlation $\langle \eta_i(t)\, \eta_j(t')\rangle = \delta_{ij}\, \delta(t-t')$, where $\delta_{ij}$ is the Kronecker delta, $\delta(.)$ is the Dirac delta function and $\langle \cdot \rangle$ represents the average over ensembles. The white noise can be naively said to be the differential of a Wiener process (\textbf{W}). 

Renaud \& Vanneste~\cite{renaud2020dispersion} have used the white-noise model for particles suspended in a TG flow. They revisited Childress' classic calculation of $\mathcal{O}(Pe^{-1/2})$ effective diffusivity of a passive scalar in a cellular flow with the inclusion of particle inertia \cite{childress1979alpha}. They showed that, in the $St\ll1$ limit, the effective diffusivity increases (decreases) for heavy (light) particles with increasing $St$. Here we follow the same approach to model the thermal diffusion of droplets in a TG flow.

The TG flow is a doubly periodic array of counter-rotating cellular vortices. The stream function for the TG flow with length scale $L_f$ and velocity scale $V_f$ is $\psi = V_f\, L_f\, \sin(x/L_f)\, \sin(y/L_f)$. The corresponding nondimensional velocity field is $\textbf{u}^* = [\sin(x^*)\, \cos(y^*), - \cos(x^*)\, \sin(y^*)]$. We use the flow length scale ($L_f$) and flow time scale ($\tau_f$) to nondimensionalise the Langevin equation Eq.~(\ref{eq1}) to get 
\begin{align} 
St \, {\rm d}\textbf{v}_p^* & = (\textbf{u}^*(\textbf{x}_p^*)-\textbf{v}_p^*)\,{\rm d}t^*+\sqrt{\frac{2}{Pe}}\, {\rm d}\textbf{W}^*~, \text{with}\label{eq3}\\
{\rm d} \mathbf{x}_p^* & = \mathbf{v}_p^* \, {\rm d}t^*, \nonumber
\end{align}
where `$*$' indicates that the parameters are nondimensional. Hereafter, we only deal with nondimensional quantities and drop the `$*$'. The Stokes number is $St = \tau_p/\tau_f = 2\, r^2\, \rho/(9\, \mu_f\, \tau_f)$ and the Péclet number is $Pe = \tau_d/\tau_f = V_f\, L_f\, 6\, \pi\, \mu_f\, r/(k_B\, T)$. The first term on the right hand side of Eq.~(\ref{eq3}) is the `drift term' and the second one is the `diffusion term'. Note that $St$ and $Pe$ are particular for each droplet, and for growing droplets, they increase with time. The parabolic growth model given by Eq.~(\ref{eq2}) can be used to obtain their instantaneous values as 
\begin{align} 
St &= St_0+\Pi\, t~,\label{eq4}\\
Pe &= Pe_0\, \sqrt{1+(\Pi/St_0)\, t}~,\label{eq5}
\end{align}
where $\Pi = \tau_p/\tau_c = 4\, \rho\, \xi/(9\, \mu_f)$ is the nondimensional droplet growth-rate, $\tau_c = r^2/(2\, \xi)$ is the condensation time scale while $St_0 = 2\, r_0^2\, \rho/(9\, \mu_f\, \tau_f)$, $Pe_0 = V_f\, L_f\, 6\, \pi\, \mu_f\, r_0/(k_B\, T)$ are the Stokes number and Péclet number based on initial droplet size.

The temperature and pressure of the atmosphere at the approximate height where cumulus clouds form are $T \approx 0^{\circ}$ C and $P \approx 80$ kPa. This yields $\xi_1 \approx 68.2 \, \mu m^2/s$ \cite{yau1996short} for water droplets. The typical supersaturation in a cloud is $s \approx 0.5 \%$. Thus the estimated value of the growth rate $\Pi$ for water droplets is around $1.4 \times 10^{-5}$. Typical Kolmogorov scales for a cloud are $L_{\eta} = 0.8$mm, $\tau_{\eta} = 0.04$s and $V_{\eta} = 2$cm/s \cite{grabowski2013growth}. Using these scales, the initial Stokes number and Péclet number for $5 \, \mu m$ water droplets are $St_0 \approx 8.15\times 10^{-3}$ and $Pe_0 \approx 6.84\times 10^{6}$. We study the dynamics for wider ranges of $St_0, Pe_0$ and  $\Pi$ than are typical in clouds, in order to better understand the effects of particle inertia, diffusion and growth.

To study the dynamics of the droplets, we integrate Eqs.~(\ref{eq3}) in time for each droplet. Since the droplets are initially micron sized, they have a small Stokes number at initial times. Eq.~(\ref{eq3}) is singular in the limit of $St \ll 1$ and its overdamped form \cite{renaud2020dispersion},
\begin{equation} 
{\rm d} \textbf{x}_p = \left(\textbf{u}-St\, \frac{{\rm D} \textbf{u}}{{\rm D} t}\right)\,{\rm d}t + \sqrt{\frac{2}{Pe}}\, {\rm d} \textbf{W}~,
\label{eq6}
\end{equation}
where ${\rm D}/{\rm D} t = \partial/\partial t +\textbf{u} \bm{\cdot \nabla}$ represents the material derivative, may be used instead. Eq.~(\ref{eq6}) is valid in the limit of $St \ll 1$, $Pe \gg 1$ and $St \cdot Pe = \mathcal{O}(1)$. While we refer to Eq.~(\ref{eq6}) to aid in understanding, the results presented here are obtained by integrating Eqs.~(\ref{eq3}) directly with small enough time steps.

We study numerically the dispersion and clustering of identical droplets randomly distributed over a selected region of the TG flow. As time progresses, the droplets get advected and diffused by the flow and the thermal noise respectively, during which they may also grow in size by condensation. The instantaneous Stokes number and Péclet number are calculated as per Eqs.~(\ref{eq4}) and (\ref{eq5}). A fourth-order Runge-Kutta scheme (RK4) is used to integrate the deterministic cases ($Pe^{-1} = 0$) of Eqs.~(\ref{eq3}), while the Euler–Maruyama method is used to integrate the stochastic cases ($Pe^{-1} \neq 0$) of Eqs.~(\ref{eq3}). The time step for integration, ${\rm d} t \leq 0.1 \, \text{min}( St, St/\Pi,Pe)$,  is a small fraction of the relevant time scales in the problem. We validate our numerical scheme by comparing our results with those of \cite{renaud2020dispersion} (see Fig.~\ref{fig6}).

The statistics of the distribution of droplets is analysed using the time evolution of the mean-square-displacement (hereafter referred to as MSD) plots. The MSD is the ensemble average of the mean square distance each droplet covered from its respective initial location
\begin{equation}
\sigma^2(t) = \langle \,  \frac{1}{N}\,\sum_{i=1}^{N} ||\textbf{x}_i(t)-\textbf{x}_i(0)||^2 \, \rangle~.
\label{eq7}
\end{equation} 
Here, the angle brackets ($\langle\cdot\rangle$) represent an average over many realisations of the initial distribution of droplets/many realisations of the thermal noise and we use the symbol $\sigma^2$ to represent the MSD. The nature of the MSD versus time curve can reveal the behaviour of the collective motion of particles/droplets: MSD curves proportional to $t^{2}$ indicate ballistic motion, while a constant MSD indicates that the droplets have attained steady states, i.e., they are all pinned at different saddle (stagnation) points, approaching them along the attractive manifolds, as discussed below.  A measure of particles' ballistic velocity can be calculated from the expression ${\rm d}  \sqrt{\sigma^2(t)}/{\rm d}  t$ in the ballistic regime. However, if the MSD is proportional to $t$, then the particles/droplets are in diffusive motion with an effective diffusivity $D_{\text{eff}} = \frac{1}{4} \frac{{\rm d}  \sigma^2(t)}{{\rm d}  t}$ (for two-dimensional flows).

\section{\label{sec3} Revisiting the role of particle inertia ($St \neq 0$, $\Pi = 0$, $Pe^{-1} = 0$)}
Previous studies of the  dispersion of finite density inertial particles in TG flow find that the particle trajectories can be periodic or chaotic depending on the values of $St$ and $\rho / \rho_f$ \cite{crisanti1992dynamics,wang1992chaotic,maxey1986gravitational,rubin1995settling}. Here we revisit the problem of dispersion of heavy inertial particles in the TG flow (following \cite{wang1992chaotic} but considering the heavy particle limit). In the absence of thermal noise, and in the limit of large $\rho/\rho_f$ but finite $St$, Eqs.~(\ref{eq3}) simplify to
\begin{subequations}
\label{eq8-whole}
\begin{eqnarray}
   \frac{{\rm d} x}{{\rm d} t} = v_x~, \quad \frac{{\rm d}  v_x }{{\rm d} t} = \frac{-v_x+\sin(x)\, \cos(y)}{St}~,\\
 \frac{{\rm d} y}{{\rm d}  t} = v_y~, \quad \frac{{\rm d}  v_y}{{\rm d}  t} = \frac{-v_y -\sin(y)\, \cos(x)}{St}~.
\end{eqnarray}
\end{subequations}
\begin{figure}[t]
\includegraphics[width =\linewidth]{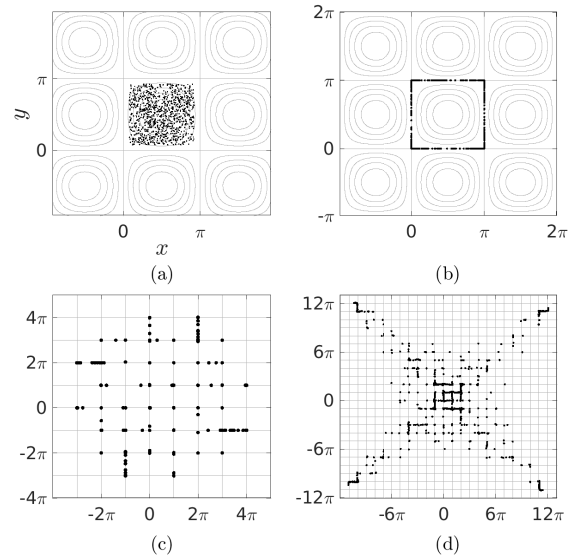}
\caption{\label{fig1}
Dispersion of $10^3$ identical inertial particles in a TG flow. The Stokes number is set at the beginning of each simulation (a) A representative initial random distribution of particles within the cell $0<x<\pi$ and  $0<y<\pi$, with $v_x = v_y =0$ at $t=0$. (b) $St = 0.1$ particles at $t = 100$, (c) $St = 0.5$ particles at $t = 100$ and (d) $St = 1.15 $ particles at $t = 100$. Note that the axes in (b-d) have different scales. }
\end{figure}
\begin{figure}[t]
\includegraphics[width = \linewidth]{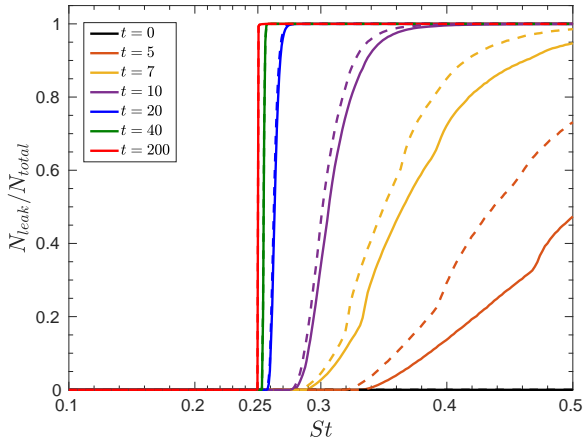}
\caption{\label{fig2} The fraction of particles that leave the TG vortex cell in which they start, plotted for various times as a function of the particle Stokes number $St$. Continuous lines represent particles starting with zero initial velocity, and dashed ones represent particles starting with the local fluid velocity. The curves are plotted using an ensemble average over $500$ simulations with $N_{total} = 10^3$ particles each. The critical Stokes number $St_{\text{cr}}$ is known to be $1/4$.}
\end{figure}
Particles are initially distributed randomly within one TG vortex cell $(0<x<\pi$, $0<y<\pi)$ as shown in Fig.~\ref{fig1}(a). The initial velocity of the particles is either set to zero or set to the local fluid velocity, with similar results. All particles with $St>0$ are centrifuged away from the vortex centre at $(\pi/2,\pi/2)$ and spiral outwards. Whether these particles remain within, or leave, the cell depends on $St$. In the long-time limit, particles with $St<St_{\text{cr}}$ remain within the area bounded by the the separatrices  $x = 0$, $x = \pi$, $y = 0$ and $y = \pi$ (see Fig.~\ref{fig1}(b)) whereas particles with $St>St_{\text{cr}}$ leave the cell. We also see that particles with $St<St_{\text{cr}}$ are ultimately absorbed by the stagnation points (hereafter referred to as SPs) at the corners of the cell (at later times than those shown here). 

The critical Stokes number $St_{\text{cr}} = 1/4$ is identified by plotting the fraction of particles that exit the initial cell (see Fig.~\ref{fig2}). This critical $St$ has been previously reported by \cite{massot2007eulerian,de2007evaluation} ($St_{\text{cr}}$ was $1/(8\, \pi)$ in their analysis due to a different choice of scaling). We calculate the leak fraction as the fraction of particles that cross the separatrices of the initial cell. Some of the particles that leave the initial cell, we note, may eventually be captured by SPs other than those of the initial cell (see Fig.~\ref{fig1}(c)). However, when $St \gtrsim 0.77$, a fraction of the particles move outwards forever with a mean direction parallel to the diagonals of the initial cell, continually crossing TG vortex cells (Fig.~\ref{fig1}(d)), and exhibiting periodic motion on open zig-zag trajectories. Similar `ballistic' motion in which the MSD scales quadratically with time has previously has been observed for inertial particles with finite density ratios in TG flow  \cite{crisanti1990passive,wang1992chaotic}.

In \cite{massot2007eulerian}, $St=1/4$ was identified as the critical Stokes number of escape across the separatrices. Below we use linear stability analysis at the SPs to describe the change in behaviour across this Stokes number, and to explain the leakage of particles to neighbouring cells for $St > 1/4$.
\subsection{Stability properties of inertial particles in TG flow}

\begin{figure}[b]
\includegraphics[width = \linewidth]{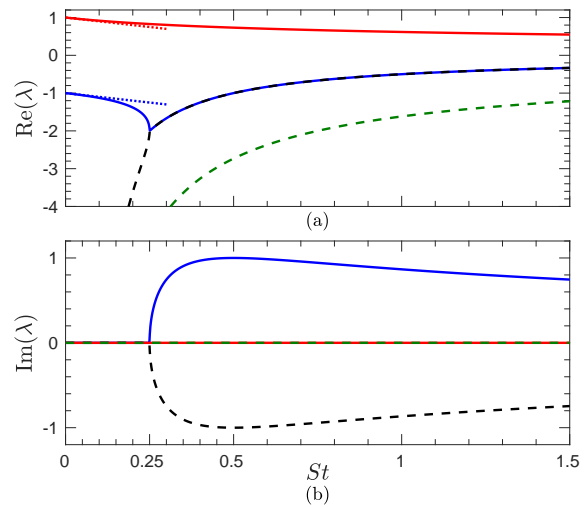}
\caption{\label{fig3}
The (a) real and (b) imaginary parts of the four eigenvalues ($\lambda$) of the linearised dynamics at a stagnation point, plotted as a function of the particle Stokes number $St$. The dotted lines represent asymptotes to eigenvalues for small $St$ obtained from the slow manifold approximation, Eq.~(\ref{eq6}).}
\end{figure}
Eqs.~(\ref{eq8-whole}) constitute a dynamical system with four variables ($x,y,v_x,v_y$). The fixed points of the system are vortex centers ($(n+1/2)\, \pi, (m+1/2)\, \pi, 0, 0$) and SPs ($n\, \pi, m\, \pi, 0, 0$) where $n, m \in \mathbb{Z}$. The system is linearised about the fixed points, with perturbations $(x',y',v_x',v_y') = (\hat{x},\hat{y},\hat{v}_x,\hat{v}_y)\, e^{\lambda\, t}$ and solved for the eigenvalues $\lambda$ to obtain exponential stability characteristics, where ($\hat{x},\hat{y},\hat{v}_x,\hat{v}_y$) are perturbation amplitudes. At the vortex centres, the eigenvalues all have positive real parts and the vortex centres behave as unstable spirals for any finite $St$ particle, explaining why particles are centrifuged away from the centre ($\pi/2,\pi/2$) in the simulations. 

\begin{figure}[t]
\includegraphics[width = \linewidth]{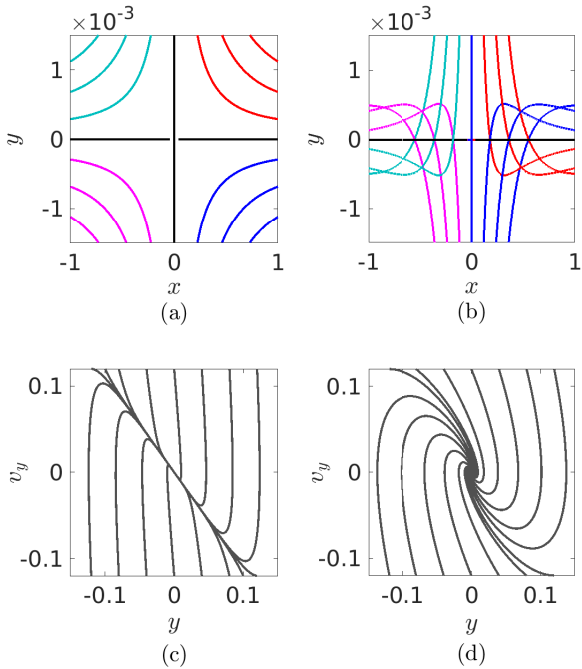}
\caption{\label{fig4} The projections of phase-space trajectories onto $x-y$ (a,b) and $y-v_y$ (c,d) planes near an SP, here placed at the origin, for particles with $St = 0.1$ (a,c) and $St = 0.5$ (b,d). Here $n+m$ is even, the axes represent separatrices in the flow, and each curve shows the trajectory of a particle. The spiralling of trajectories in the phase plane in (d) is a signature of the crossing of the flow separatrices by the particles.}
\end{figure}

The behaviour at the SPs is more complicated and the eigenvalues are plotted as a function of the Stokes number in Fig.~\ref{fig3}. For $St < 1/4$, all the eigenvalues are purely real, and one of them is positive. Such a fixed point is termed a `$3:1$ saddle' \cite{hofmann2018visualization}. For $St >1/4$, two of the eigenvalues become complex conjugates, while the positive eigenvalue remains positive; the fixed point is thus a `spiral-$3:1$ saddle' \cite{hofmann2018visualization}. The change in the four-dimensional phase space behaviour is best shown in the two-dimensional projections in Fig.~\ref{fig4}. The trajectories of particles of $St < 1/4$, asymptote to the separatrices and do not cross them (see Fig.~\ref{fig4}(a)), whereas particles of $St > 1/4$ can cross separatrices. In phase space the latter support spiral trajectories at the Stokes number shown, which is consistent with the ability to cross separatrices in finite time (see Fig.~\ref{fig4}(b)).

The existence of a positive eigenvalue indicates that the SPs are linearly unstable fixed points for inertial particles of finite $St$. However, the phase-space behaviour of these unstable fixed points changes when the $St$ exceeds $1/4$, and the phase space trajectories attain a spiral nature as well. While this change to unstable spiral-saddle behaviour does not explain why particles with $St<1/4$ remain inside the initial cell (see Fig.~\ref{fig2}), we expect that the agreement between the $St_{\text{cr}}$ found numerically and from the linear stability analysis here is not simply coincidental. In fact the connection can be clearly explained, as done in the following subsection.

\subsection{\label{subsec3b} The threshold to cross a separatrix }
Numerical simulations of Eqs.~(\ref{eq8-whole}) reveal that particles with $St > 1/4$ always cross the separatrices in the vicinity of one of the SPs. Therefore, we linearise Eqs.~(\ref{eq8-whole}) at a general SP ($n\, \pi, m\, \pi, 0, 0$) where $n, m \in \mathbb{Z}$, to get
\begin{subequations}
\label{eq9-whole}
\begin{eqnarray}
   \frac{{\rm d} x'}{{\rm d} t} = v_x'~, \quad  \frac{{\rm d} v_x' }{{\rm d} t} = \frac{-v_x'+(-1)^{n+m}\, x'}{St}~,\\
 \frac{{\rm d} y'}{{\rm d} t} = v_y'~, \quad \frac{{\rm d} v_y'}{{\rm d} t} = \frac{-v_y' -(-1)^{n+m}\, y'}{St}~.
\end{eqnarray}
\end{subequations}
where $(x',y',v_x',v_y')$ are perturbation quantities. Since the $x$ and $y$ equations are decoupled, we can combine them and rewrite Eqs.~(\ref{eq9-whole}) as follows
\begin{eqnarray}
   St\, \frac{{\rm d}^2 x'}{{\rm d} t^{2}} + \frac{{\rm d} x'}{{\rm d} t}-(-1)^{n+m}\, x' = 0~, \label{eq10} \\
   St\, \frac{{\rm d}^2 y'}{{\rm d} t^{2}} + \frac{{\rm d} y'}{{\rm d} t}+(-1)^{n+m}\, y' = 0~, \label{eq11}
\end{eqnarray}
which are the equations for damped harmonic oscillators with two degrees of freedom. The two oscillators have opposite stability, since they have oppositely signed stiffness coefficients (i.e., if $n+m$ is an even integer, then the $x-$oscillator is unstable while the $y-$oscillator is stable, and vice-versa). Here, without loss of generality, we consider SPs with even $n+m$ to explain things unless otherwise specified. The behaviour at SPs with odd $n+m$ are obtained by exchanging $x$ and $y$.

Eq.~(\ref{eq11}), therefore, represents a damped harmonic oscillator in the $y$ direction, with a positive stiffness coefficient. The damping coefficient for the system is $1$, and the critical damping factor is $2\, \sqrt{St}$, giving a damping ratio of $1/\sqrt{4\, St}$. Therefore, the system is overdamped for $St<1/4$, and underdamped for $St>1/4$. The oscillations in $y'$ for $St>1/4$ are about the horizontal separatrix connected to the SP, and thus the particle crosses the separatrix in the $y$ direction near the SP. For SPs with $n+m$ is odd, the identical argument reads: oscillations in $x'$ for $St>1/4$ about vertical separatrices makes the particle to cross separatrices in $x$ direction near SPs. Since this argument is true at all SPs, we conclude that particles can only cross the separatrices if $St>1/4$.

Eqs.~(\ref{eq10}) and (\ref{eq11}) are, in fact, exactly solvable. For an initial condition ($x_0',y_0',v_{x0}',v_{y0}'$), and $n+m$ even, the exact solutions are
\begin{eqnarray}
   x' =  C_1\, e^{\frac{(\alpha-1)\, t}{2\, St} }+
   C_2\, e^{\frac{-(\alpha+1)\, t}{2\, St} } ~, \label{eq12}\\
    y' =  C_3\, e^{\frac{(\beta-1)\, t}{2\, St} }+
   C_4\, e^{\frac{-(\beta+1)\, t}{2\, St} } ~, \label{eq13}
\end{eqnarray}
where $\alpha = \sqrt{1+ 4\, St}$, $\beta = \sqrt{1- 4\, St}$, $C_1 = x_0'\, (1+\alpha)/(2\, \alpha)+v_{x0}'\, St/\alpha$, $C_2 = x_0'\, (-1+\alpha)/(2\, \alpha)-v_{x0}'\,St/\alpha$, $C_3 = y_0'\,(1+\beta)/(2\, \beta)+v_{y0}'\,St/\beta$ and $C_4 = y_0'\,(-1+\beta)/(2\, \beta)-v_{y0}'\,St/\beta$. 

From these solutions, it can be seen that the nature of the system changes at $St_{\text{cr}} = 1/4$. Furthermore, the time taken by a particle with $St > 1/4$ to cross the horizontal separatrix $y=m \, \pi$ and escape the cell (escape time, $t_{esc}$) is the smallest of the solution for $y'(t_{esc}) = 0$, and is
\begin{eqnarray}
t_{esc} \sim \frac{2\, St}{\sqrt{4\, St - 1}}\, \left\{\pi-\tan^{-1}\left( \frac{\sqrt{4\, St-1}}{1+2\, St\, \frac{v_{y0}'}{y_0'}}\right) \right\}~.
\label{eq14}
\end{eqnarray}
By this time, the particle could typically be sufficiently far away from the SP in the $x$ direction so that the linearised system no longer governs further dynamics. Thus, Eq.~(\ref{eq14}) would be only a rough estimate for the escape time of the particles in the $y$ direction across horizontal separatrices. 

From the exact solution Eq.~(\ref{eq12}), we also see that particles with a sufficiently large initial velocity, directed specifically, can cross the vertical separatrices as well. The magnitude of critical velocity can be obtained from  Eq.~(\ref{eq12}) as $|v_{x0}'| > v_{\text{cr}} = 2\,|x_0'| /(-1+\sqrt{1+4\, St})$, and should be directed towards the vertical separatrix $x = n\, \pi$. For $St \gtrsim 1/4$, the particles can usually have that much velocity; however, that will not be directed towards the vertical separatrix, instead directed away from it near any SP, due to the centrifuging effect of the vortex. Extra forces in the system like gravity, acting towards the vertical separatrix could activate this criterion. Thus, it is not relevant in explaining the leakage of particles with $St \gtrsim 1/4 $ from the initial TG vortex cell in the present system.

\begin{figure}[t]
\includegraphics[width =\linewidth]{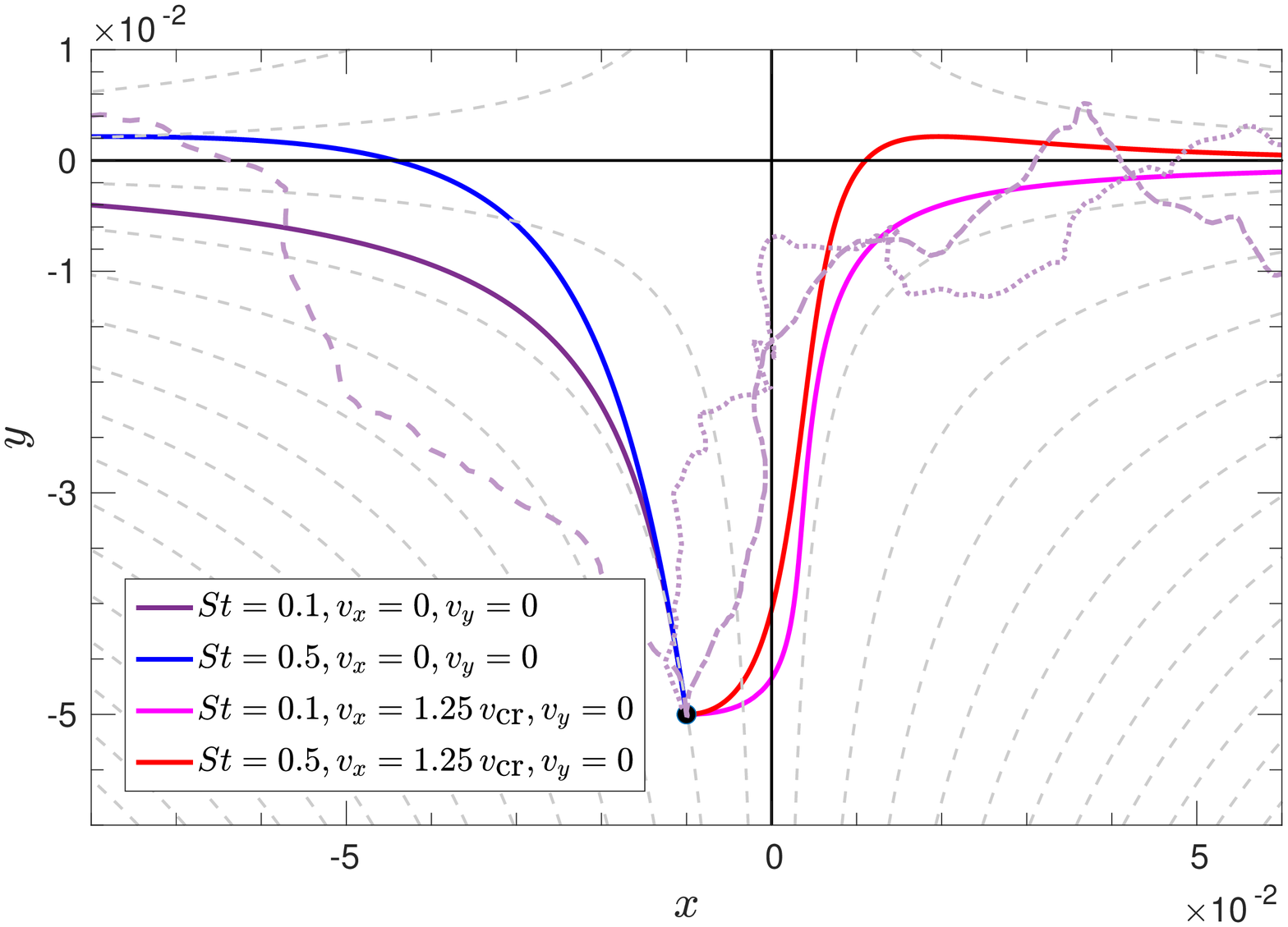}
\caption{\label{fig5} The trajectories of inertial particles with different Stokes numbers $St$ and different initial velocities $v_x$ (continuous lines) starting from the same location near the SP at the origin in a TG flow represented by the black dot ($-0.01,-0.05$). These trajectories could cross the horizontal and vertical separatrices depending on the initial conditions. Faded purple dashed or dotted lines represent trajectories of $St = 0.1$ particles with zero initial velocity, for different realisations of thermal noise ($Pe = 10^4$), and show that even $St<1/4$ particles could cross the separatrices for individual realisations of the noise. The thin dashed grey lines are the streamlines of the flow.}
\end{figure}

In Fig.~\ref{fig5}, we plot the trajectories of inertial particles starting near an SP placed at the origin ($n+m=0$). The axes coincide with separatrices. When the initial velocity is large, the trajectories cross the $x=0$ separatrix; when the initial Stokes number is large, the trajectories cross the $y=0$ separatrix; when both the initial velocity and the Stokes number are large, trajectories cross both the $x=0$ and the $y=0$ separatrices. Examples are shown in the figure.

\section{\label{sec4} The role of thermal noise \\($St \neq 0$, $\Pi = 0$, $Pe^{-1} \neq 0$) }
\begin{figure}[b]
\includegraphics[width = \linewidth]{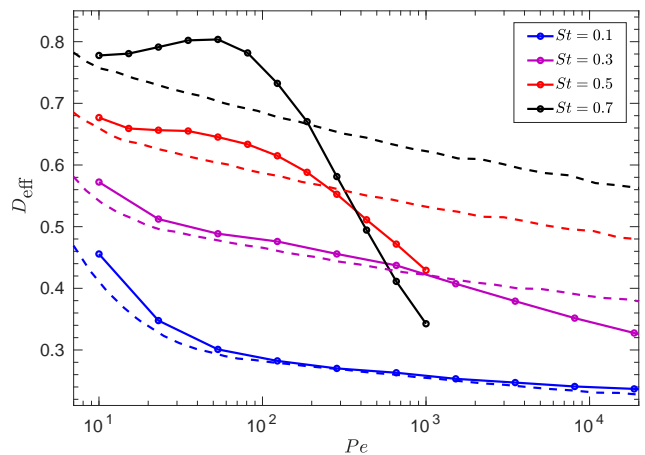}
\caption{\label{fig6} Effective diffusivity for heavy particles in TG flow against Péclet number $Pe$ at various Stokes numbers. Our results (solid lines with markers) are compared with the expression of Renaud \& Vanneste~\cite{renaud2020dispersion} (dashed lines).}

\end{figure}

In the weak molecular diffusion limit ($Pe\gg1$), the effective diffusivity of passive scalars crucially depends on the flow topology. In shear flows with open streamlines $D_{\textrm{eff}}\sim Pe$ - the classical Taylor-Aris dispersion \cite{taylor1953dispersion,aris1956dispersion,young1991shear}. For cellular flows, molecular diffusion becomes dominant in a thin boundary layer near the separatrices, assisting migration across cells, leading to $D_{\textrm{eff}}\sim Pe^{-1/2}$ \cite{childress1979alpha}. To understand the enhanced transport due to convection in the above two scenarios, one should recall that the diffusivity in the absence of flow is $D\sim Pe^{-1}$. The effective diffusivity $D_{\text{eff}}$ of inertial particles in a TG flow, with the asumption of $St\ll1$, $Pe\gg1$ and $St \, Pe=\mathcal{O}(1)$, was recently calculated by \cite{renaud2020dispersion}. For $St = 0.1$, our simulated results find excellent agreement with theirs, as shown in Fig.~\ref{fig6}. At higher values of $St$, however, the expression of \cite{renaud2020dispersion} is no longer accurate. As $St$ increases, we find that the effective diffusivity acquires a non monotonic variation with $Pe$, and decreases rapidly for large $Pe$, in qualitative departure from the distinguished limit of $St\,Pe = O(1)$.
The diffusion of inertial particles in periodic, shear, and elongational flows are studied in  \cite{san1979brownian,rubi1988brownian,fannjiang1994convection,subramanian2002inertial}. The study by Rubi \& Bedeaux~\cite{rubi1988brownian} on elongational flows is of particular interest since, near the SPs, TG flow resembles elongational flow. The linearised governing equations near an SP ($n\, \pi, m\, \pi, 0, 0$) when $Pe^{-1}>0$ read
\begin{eqnarray}
   St\, \frac{{\rm d}^2 x'}{{\rm d} t^{2}} + \frac{{\rm d} x'}{{\rm d} t}- (-1)^{n+m}\, x' = \sqrt{\frac{2}{Pe}}\, \eta_x(t)~, \\
   St\, \frac{{\rm d}^2 y'}{{\rm d} t^{2}} + \frac{{\rm d} y'}{{\rm d} t}+ (-1)^{n+m} \, y' = \sqrt{\frac{2}{Pe}}\, \eta_y(t)~.
\end{eqnarray}
The MSD of a particle near the elongational flow can be calculated as (for $n+m$ even)
\begin{eqnarray}
   \langle x'^2 \rangle =C_1^2\, e^{\frac{(\alpha-1)\, t}{St}}+2\, C_1\, C_2\, e^{\frac{-t}{St}} + C_2^2\, e^{\frac{-(\alpha+1)\, t}{St}} \nonumber \\
   \frac{1}{\alpha^2\, Pe}\, \bigg\{e^{\frac{-t}{St}}\, \left[\cosh\left(\frac{\alpha\, t}{St}\right)+\alpha\, \sinh\left(\frac{\alpha\, t}{St}\right)\right] \nonumber \\
   -1-4\, St\, (1-e^{\frac{-t}{St}})  \bigg\}~,\label{eq:disp_x}\\
   \langle y'^2 \rangle =C_3^2\, e^{\frac{(\beta-1)\, t}{St}}+2\, C_3\, C_4\, e^{\frac{-t}{St}} + C_4^2\, e^{\frac{-(\beta+1)\, t}{St}} \nonumber \\
   \frac{1}{\beta^2\, Pe}\, \bigg\{-e^{\frac{-t}{St}}\, \left[\cosh\left(\frac{\beta\, t}{St}\right)+\beta\, \sinh\left(\frac{\beta\, t}{St}\right)\right] \nonumber \\
   +1-4\, St\, (1-e^{\frac{-t}{St}})  \bigg\} ~, \label{eq:disp_y}
\end{eqnarray}
where $\alpha, \beta$ and the constants $C_1, C_2, C_3$ and $C_4$ are as defined in section \ref{sec3}. Eqs.~(\ref{eq:disp_x}) and (\ref{eq:disp_y}) generalise the expressions in \cite{rubi1988brownian} to arbitrary initial conditions and $St$. At short times, the MSD scales as $\langle x'^2 + y'^2 \rangle\sim 4\, t^{3}/(3\, Pe\, St^2) $. Thus, when $Pe^{-1}>0$, particles of any $St$ can cross separatrices and escape the initial cell. This tendency increases for stronger noise (smaller $Pe$). Fig.~\ref{fig5} shows sample trajectories of  particles with $Pe=10^4$ and zero initial velocity. For the same initial condition, individual realisations of the thermal noise may lead to trajectories crossing the separatrices.
\section{\label{sec5} The role of condensation\\ ($St \neq 0$, $\Pi \neq 0$, $Pe^{-1} = 0$)}
We next study the dynamics of growing (by condensation) droplets in the TG flow without thermal noise. Droplets of initial Stokes number $St_0 = 0.1$ and zero initial velocity are distributed randomly in a square patch as shown in Fig.~\ref{fig1}(a), and allowed to grow with a growth-rate $\Pi = 10^{-2}$. We note that a different choice of initial velocity (the local fluid velocity) does not change the dynamics qualitatively. We solve Eqs.~(\ref{eq3}) with $Pe^{-1}\equiv 0$. The advective motion by the flow dominates the initial dynamics of the droplets, where the droplets are thrown out of the vortex center ($\pi/2,\pi/2$). As time progresses, the instantaneous Stokes number of droplets, $St(t) = St_0+\Pi\, t$ exceeds $1/4$, allowing droplets to cross the separatrices and spread in a manner qualitatively similar to that seen in Fig.~\ref{fig1}(c). Unlike fixed $St$ particles, all continuously growing droplets eventually enter the ballistic regime of motion (qualitatively as will be seen in Fig.~\ref{fig9}(d)). In this phase, the droplets are observed to travel along $45^\circ-135^\circ$ lines in a zig-zag manner, which is similar to the open-trajectory periodic motion identified in Wang et al.~\cite{wang1992chaotic}. However, for sufficiently small growth rates ($\Pi \lesssim 0.005$), our simulations show that growing droplets get trapped at the stagnation points instead of attaining ballistic velocities. Once these droplets are trapped (to within numerical precision) at the SPs, their velocities remain zero despite their continuous growth in size.

\begin{figure}[b]
\includegraphics[width = \linewidth]{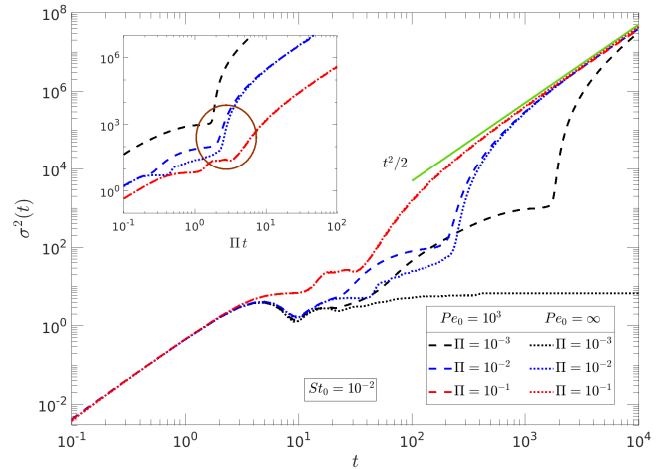}
\caption{\label{fig7} MSD for $St_0 = 10^{-2}$ droplets with ($Pe_0 = 10^3$) and without ($Pe_0 = \infty$) thermal noise, for different growth rates, plotted against time. The asymptote for the long time ballistic motion is shown in green. The circle within the inset indicates that the switch to the ballistic regime occurs at a scaled time $\Pi\, t \sim O(1)$.}
\end{figure}
\begin{figure}[b]
\includegraphics[width = \linewidth]{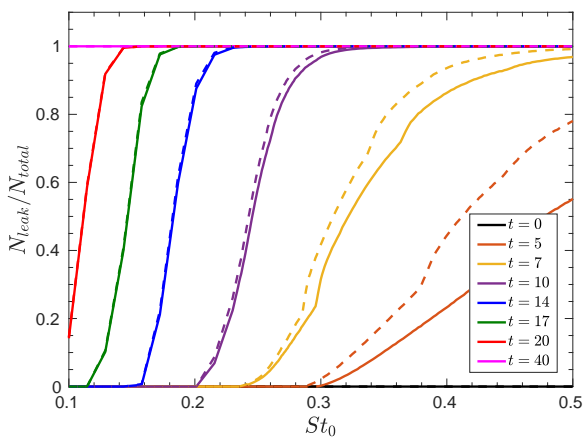}
\caption{\label{fig8} The fraction of condensing droplets ($\Pi = 10^{-2}$) that exit their initial cell is plotted against $St_0$ for various simulation times. Continuous lines indicate the case of particles with zero initial velocity, and dashed ones represent particles initially at the local fluid velocity. Simulations are performed with $N_{total} = 10^3$ particles over $500$ realisations.}
\end{figure}

We plot the MSD of the droplets, ensemble-averaged over $10^3$ realisations of initial distributions of $10^3$ particles each, for an initial Stokes number $St = 10^{-2}$, three different values of growth-rates ($\Pi = 10^{-3}, 10^{-2}$ and $10^{-1}$) in Fig.~\ref{fig7} ($Pe_0 = \infty$ case).  Initially, i.e., for $t\ll 1$, the MSD grows independent of $\Pi$. In the intermediate phase ($t \sim \mathcal{O}(10)-\mathcal{O}(10^3)$), the MSD has a clear dependence on $\Pi$. The curves in this phase display waviness, which could be caused by droplets hopping back and forth between neighbouring cells. At large times ($t > \mathcal{O}(10^3)$), the MSD scales as $t^2/2$ and droplets enter the ballistic regime. The time at which the dynamics becomes ballistic is approximately the same time at which $St(t) \sim \mathcal{O}(1)$. This time decreases for larger $\Pi$ as  $t \sim 1/\Pi$, as shown in the inset of Fig.~\ref{fig7}. The scaling $\Pi\, t$ was obtained empirically. For the lowest growth rate $\Pi= 10^{-3} < 0.005$ shown in the figure, the MSD is saturated at large times, indicating that the droplets are trapped at the SPs of the flow even though they are continuously growing.

For other values of initial Stokes number $St_0$ (not shown), the MSD plot is qualitatively the same as shown in Fig.~\ref{fig7}. We also see that for sufficiently large $\Pi$ and long times, the numerical value of the MSD becomes independent of $St_0$ and $\Pi$. Empirically we obtained that the asymptotic fit in this phase is $\sigma^2(t) \sim t^2/2$, indicating that the measure of the nondimensional ballistic velocity of droplets in this phase asymptotically reaches the value $1/\sqrt{2}$ for large $St$ (see Sec.~\ref{sec2}).

The evolution of the leakage fraction of particles against their initial Stokes number $St_0$ is plotted in Fig.~\ref{fig8}. Unlike in Fig.~\ref{fig2}, there is no critical initial Stokes number $St_0$ for growing particles, as one would intuitively expect. In the following subsection, we examine the reason using a local analysis near SPs. Also, we obtain an analytical expression for the escape time of droplets from a vortex cell.

\subsection{Local analysis near SPs}
We observe from the numerical simulations that condensing droplets, like constant-size particles, also cross the separatrices near SPs. The linearized perturbation equations (see section \ref{subsec3b}) for a droplet with a Stokes number $St_1 > St_0$ (allowing for droplet growth by the time it reaches the vicinity of an SP) near the SP $(n\, \pi, m\, \pi, \, 0, 0)$ read
\begin{eqnarray}
   (St_1+\Pi \, t) \, \frac{{\rm d}^2 x'}{{\rm d} t^{2}} + \frac{{\rm d} x'}{{\rm d} t}-(-1)^{n+m}\, x' = 0~,  \\
   (St_1 + \Pi \, t) \, \frac{{\rm d}^2 y'}{{\rm d} t^{2}} + \frac{{\rm d} y'}{{\rm d} t}+(-1)^{n+m}\, y' = 0~. 
\end{eqnarray}
The system is thus comprised of damped harmonic oscillators with two degrees of freedom, but with increasing mass. Using the transformation $t_1 = t + St_1/\Pi$, these equations for SPs with even $n+m$ can be written as
\begin{eqnarray}
   \Pi\, t_1 \, \frac{{\rm d}^2 x'}{{\rm d} t_1^{2}} + \frac{{\rm d} x'}{{\rm d} t_1}- x' = 0~,  \label{eq21}\\
   \Pi\, t_1 \, \frac{{\rm d}^2 y'}{{\rm d} t_1^{2}} + \frac{{\rm d} y'}{{\rm d} t_1}+ y' = 0~. \label{eq22}
\end{eqnarray}
The general solutions are in terms of Bessel functions,
\begin{eqnarray}   x' = t_1^{-\frac{\gamma}{2}}\, \left\{C_5\, \text{I}_{\gamma}\left(2\, \sqrt{\frac{t_1}{\Pi}}\right)+C_6\, \text{I}_{-\gamma}\left(2\, \sqrt{\frac{t_1}{\Pi}}\right) \right\}~, \label{eq23}\\
   y' = t_1^{-\frac{\gamma}{2}}\, \left\{C_7\, \text{J}_{\gamma}\left(2\, \sqrt{\frac{t_1}{\Pi}}\right)+C_8\, \text{J}_{-\gamma}\left(2\, \sqrt{\frac{t_1}{\Pi}}\right) \right\}~, \label{eq24}
\end{eqnarray}
where $\gamma = -1+\Pi^{-1}$. Since these expressions are not particularly helpful in the limit ($\Pi \rightarrow 0$), because of the singular nature of arguments of Bessel functions, we use the Wentzel–Kramers–Brillouin (WKB) method to obtain the asymptotic solutions for $\Pi \rightarrow 0$ (see Appendix \ref{appA})
\begin{eqnarray}
   x' \sim \frac{t_1^\frac{-1}{2\, \pi}\, \left\{C_9\, \exp(-\chi_{+})+C_{10}\, \exp(\chi_{+}) \right\}}{\sqrt[4]{\Pi\, \left(\frac{1}{t_1}-\frac{1}{2\, t_1^2}\right)+\frac{1}{4\, t_1^2}}} ~, \label{eq25}\\
      y' \sim \frac{t_1^\frac{-1}{2\, \pi}\, \left\{C_{11}\, \sin \chi_{-}+C_{12}\, \cos \chi_{-} \right\}}{\sqrt[4]{\Pi\, \left(\frac{1}{t_1}+\frac{1}{2\, t_1^2}\right)-\frac{1}{4\, t_1^2}}}~, \label{eq26}
\end{eqnarray}
where 
\begin{eqnarray}
   \chi_{\pm} = \frac{1}{\Pi}\, \int_{\frac{St_1}{\Pi}}^{t_1}\sqrt{\Pi\, \left(\frac{1}{\tau} \mp  \frac{1}{2\, \tau^2}\right) \pm \frac{1}{4\, \tau^2}}\, {\rm d}\tau~. \label{eq27}
\end{eqnarray}
The lower limit of the integral is taken as the value of $t_1$ corresponding to $t=0$, i.e. $St_1/\Pi$. The constants $C_5 ... C_{12}$ depend on the initial conditions of the perturbation ($x_0',y_0',v_{x0}',v_{y0}'$). The form of the asymptotic solution Eq.~(\ref{eq26}) implies that there exists a critical Stokes number $(1-2\, \Pi)/4$, a modification to $St_{\text{cr}} = 1/4$ of fixed $St$ particles. When $St_1 > (1-2\, \Pi)/4$, the behaviour of $y'$ would be oscillatory, similar to the case of a fixed $St$ particle, but the time-variation in $St$ is accounted for here. In contrast, when $St_1 < (1-2\, \Pi)/4$ the scenario is different from that for constant $St$. At a time $t_{\text{TP}} = (1-2\, \Pi-4\, St_1)/(4\, \Pi)$ there is a `turning point' by WKB analysis (see Appendix \ref{appA}), close to which the WKB solution Eq.~(\ref{eq26}) will not be valid. However, when $t \gg t_{\text{TP}}$, this oscilatory solution will be valid even for $St_1 < (1-2\, \Pi)/4$, which could allow the droplet to cross the separatrices. Again the time taken to cross a horizontal separatrix $y=m\, \pi$ would be the smallest of the solution of $y'(t_{esc}) = 0$. Using Eq.~(\ref{eq24}), the actual estimate would be the solution $t_{esc}$ of the following transcendental equation
\begin{eqnarray}
      \frac{\text{J}_{\gamma}\left(\frac{2}{\sqrt{\Pi}}\, \sqrt{t_{esc}+\frac{St_1}{\Pi}} \right)}{\text{J}_{-\gamma}\left(\frac{2}{\sqrt{\Pi}}\, \sqrt{t_{esc}+\frac{St_1}{\Pi}} \right)} = \nonumber \\ \frac{\text{J}_{1+\gamma}\left(\frac{2\,\sqrt{St_1}}{\Pi} \right) + \frac{v_{y0}'}{y_0'}\, \sqrt{St_1}\, \text{J}_{\gamma}\left(\frac{2\,\sqrt{St_1}}{\Pi} \right)}{-\text{J}_{-1-\gamma}\left(\frac{2\,\sqrt{St_1}}{\Pi} \right) + \frac{v_{y0}'}{y_0'}\, \sqrt{St_1}\, \text{J}_{-\gamma}\left(\frac{2\,\sqrt{St_1}}{\Pi} \right)}~.
\end{eqnarray}
An approximate estimate can be obtained using the WKB solution Eq.~(\ref{eq26}) as $t_{esc} \sim t_{\text{TP}} + (9\, \pi/ (32\, \sqrt{\Pi}))^{2/3}$ when $St_1 < (1-2\, \Pi)/4$ (see Appendix \ref{appB}), indicating that as $\Pi$ decreases, the escape time increases. A more accurate expression using WKB is given in the Appendix \ref{appB}, both for $St_1 < (1-2\, \Pi)/4$ and $St_1 > (1-2\, \Pi)/4$ cases. 

As we mentioned earlier, for the case of non-condensing particles, this exit time is a rough estimate from the linear theory. By this time, the droplet could be sufficiently away from SP so that nonlinear effects could alter this exit time.
\section{\label{sec6} Combined effects of condensation and thermal noise \\($St \neq 0$, $\Pi \neq 0$, $Pe^{-1} \neq 0$)}
We now study the dynamics of condensing droplets in TG flow with thermal noise by solving the full stochastic Langevin equation Eq.~(\ref{eq3}). Since the droplets are condensing, both $St$ and $Pe$ increases with time as per Eqs.~(\ref{eq4}) and (\ref{eq5}). The strength of the thermal noise is inversely proportional to the $Pe$. Thus, as time progresses, the influence of thermal noise becomes weaker. Diffusive behaviour takes a long time to be set up even for droplets which are not growing, i.e., when the thermal noise is not decreasing in strength with time. It would take even longer for growing droplets. Similarly, ballistic motion would take less time to be set up for growing droplets. We therefore expect that at long times, ballistic dynamics will be predominant. The intermediate time behaviour, where the effects of advection and thermal diffusion may be in competition, is not easy to anticipate.
\begin{figure}[t]
\includegraphics[width =\linewidth]{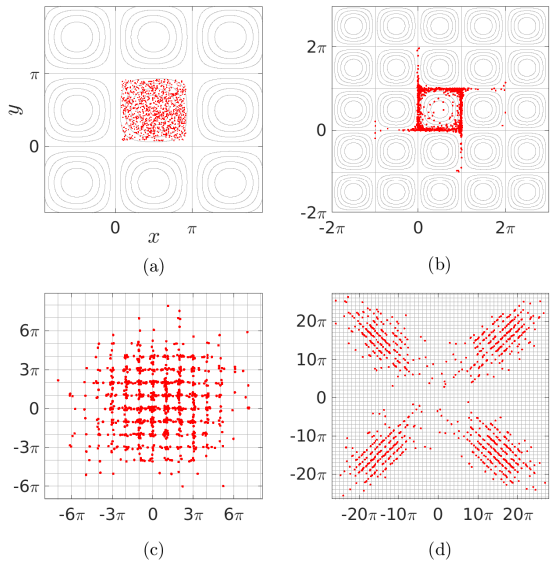}
\caption{\label{fig9}
Dispersion of identical condensing droplets ($N = 10^3, \, St_0 = 0.1, \, \Pi = 10^{-2}$) in TG flow with thermal noise ($Pe_0 = 10^3$) at  (a) $t = 0$, (b) $t = 15$, (c) $t = 150$ and (d) $t = 300$. Initial velocity of all the droplets are chosen to be zero.}
\end{figure}

We numerically study the dynamics of initially identical droplets ($St_0 = 0.1$ and $Pe_0 = 10^3$, $\mathbf{v}=0$ distributed randomly over the TG vortex cell, as shown in Fig.~\ref{fig9}(a). The growth rate is chosen as $\Pi = 10^{-2}$. For short times, advection and stochastic forcing together make the particles to cross separatrices despite having Stokes numbers $St < 1/4$ (see Fig.~\ref{fig9}(b)). For larger times, the combined effects of condensation growth and thermal noise lead to the greater diffusion than with condensation alone (see Fig.~\ref{fig9}(c)). As expected, in large time limit, the droplets move ballistically along $45^\circ-135^\circ$ paths (see Fig.~\ref{fig9}(d)).

The MSD, ensemble-averaged over $10^3$ realisations with $10^3$ droplets each with $St_0 =10^{-2}$ and $Pe_0=10^3$, is plotted in Fig.~\ref{fig7} for three different values of the growth rate $\Pi$. We see that thermal noise (finite $Pe$) significantly alters the dynamics for small droplet growth rates $\Pi = 10^{-3}$, leading to ballistic motion instead of droplets trapped at SPs.

\begin{figure}[t]
\includegraphics[width =\linewidth]{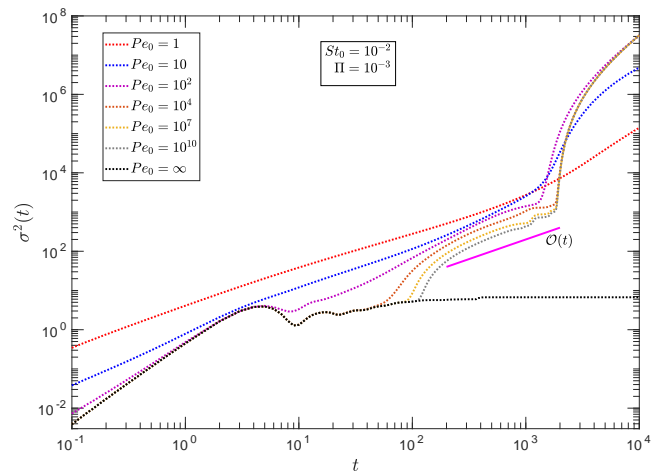}
\caption{\label{fig10} MSD for droplets with $St_0 = 10^{-2}$ and $\Pi =10^{-3}$, for different $Pe_0$. The $\mathcal{O}(t)$ asymptote, indicative of a diffusive regime, is shown in magenta.}
\end{figure}

In Fig.~\ref{fig10}, we plot the MSD for different $Pe_0$ but the same $St_0$ and $\Pi$, showing that the intermediate phase becomes smoother for smaller $Pe_0$ (and thus larger thermal noise). Furthermore, the intermediate regime scales linearly with $t$ indicating that the behaviour is diffusive. As $Pe_0$ decreases, this diffusive regime becomes wider.

\begin{figure}[b]
\includegraphics[width = \linewidth]{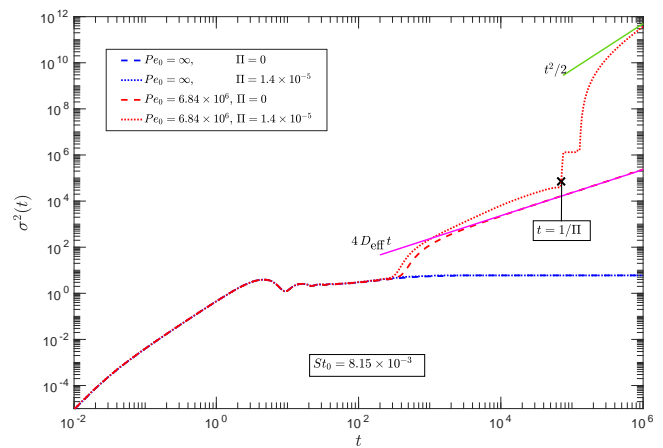}
\caption{\label{fig11} MSD versus time for $5 \, \mu m$ droplets with initial Stokes number $St_0 = 8.15 \times 10^{-3}$, realistic $\Pi$ and $Pe_0$ compared with non-condensing and non-diffusing droplets. Asymptotes for ballistic and diffusive behaviour are shown. For $\Pi=0$ and finite $Pe_0$, the diffusivity $D_{\text{eff}}=0.05786$ matches that calculated from the expression of \cite{renaud2020dispersion}. Growing droplets subject to thermal noise transition from diffusive to ballistic behaviour at $t \sim 1 / \Pi$.}
\end{figure}

Lastly, we study the behaviour of droplets with parameter values representative of atmospheric clouds: $St_0 = 8.15 \times 10^{-3}$, $Pe_0 = 6.84 \times 10^6$, $\Pi = 1.4 \times 10^{-5}$ (see Sec.~\ref{sec2}). The MSD, ensemble-averaged over $100$ realisations of $10^4$ particles each, is plotted in Fig.~\ref{fig11}. We note that fewer realisations were possible due to the extremely long times for which these simulations need to be run. For comparison, the MSD curves corresponding to the special cases studied in sections \ref{sec3}--\ref{sec5} are also plotted in Fig.~\ref{fig11}.
It is observed that after $10^6$ nondimensional time (11 hours), the enhancement in MSD is of $\mathcal{O}(10^{10})$ by the inclusion of both condensation and thermal noise. We observe that the MSD for $Pe_0 = \infty$ cases is independent of whether condensation occurs, and $\Pi > 0$ (still $\Pi < 0.005$) leads to a constant MSD indicative of a steady state. Here, this steady state is achieved due to the capture of droplets at the SPs of the TG flow, despite the fact that the droplet Stokes numbers have increased to $St=\mathcal{O}(10)$ by $t=10^6$ (cf. the discussion of figure \ref{fig7}).

With nonzero thermal noise, droplets behave diffusively both for $\Pi=0$ and $\Pi>0$, although growing droplets show departures from the asymptote ($4\, D_\text{eff}\, t$), eventually transitioning to ballistic motion at $t \sim 1/\Pi$, as expected from section \ref{sec5}. The combined effects of condensation and thermal noise are reflected in the greater diffusion at intermediate times.
\section{\label{sec7} Conclusion}
We studied the effects of thermal noise and condensation on the dispersion of monodisperse droplets suspended in a Taylor-Green vortex flow.
In the absence of thermal noise and condensation, we found, in agreement with \cite{massot2007eulerian}, that droplets with $St<1/4$ remain trapped in their initial vortices. We showed that the addition of either thermal noise or a finite condensation rate removes this condition, increasing droplet dispersion by orders of magnitude. We showed that droplets growing by condensation typically attain a state of ballistic motion away from their initial vortices for times $t\gtrsim\Pi^{-1}$, travelling along $45^\circ$ diagonal trajectories with average nondimensional velocities of $1/\sqrt{2}$, but that sufficiently small growth rates $\Pi \lesssim  0.005$ allow droplets to remain trapped at the stagnation points of the flow. We showed that, in the presence of thermal noise, this transition from the trapped state to the ballistic state occurs proceeds through an intermediate diffusive regime where the mean squared displacement of the droplets grows linear as $\sigma^2 \sim t$. Our results with this model flow are encouraging, and suggest further studies where the effects of polydispersity and collision-coalescence of droplets, their gravitational settling, and the effects of latent heating and buoyancy are included.

\begin{acknowledgments}
SR is supported through Swedish Research Council grant no. 638-2013-9243.
RG acknowledges support of the Department of Atomic Energy, Government of India, under project no. RTI4001. AVSN would like to thank the Prime Minister's Research Fellows (PMRF) scheme, Ministry of Education, Government of India. AR and AVSN would like to acknowledge the support from Laboratory for Atmospheric and Climate Sciences, Indian Institute of Technology Madras.
\end{acknowledgments}

\appendix
\section{\label{appA} WKB analysis for condensing droplets near SPs}
By eliminating first order terms, Eqs.~(\ref{eq21}) and (\ref{eq22}) can be rewritten as,
\begin{eqnarray}
   \Pi^2\,  \frac{{\rm d}^2 x''}{{\rm d} t_1^{2}} + \left\{-\frac{\Pi}{t_1}+\left(\Pi-\frac{1}{2}\right)\, \frac{1}{2\, t_1^2} \right\}\, x'' = 0~,  \label{A1}\\
    \Pi^2\,  \frac{{\rm d}^2 y''}{{\rm d} t_1^{2}} + \left\{\frac{\Pi}{t_1}+\left(\Pi-\frac{1}{2}\right)\, \frac{1}{2\, t_1^2} \right\}\, y''= 0~, \label{A2}
\end{eqnarray}
where $x'' = x'\, t_1^{1/(2\, \Pi)}$ and $y'' = y'\, t_1^{1/(2\, \Pi)}$. These equations resemble the form of differential equations
\begin{equation}
    \Pi^2\, \frac{{\rm d}^2 \phi}{{\rm d} t^{2}}+q(t)\, \phi = 0~, \label{A3}
\end{equation}
 which can be asymptotically solved using the WKB method \cite{bender1978sa}. For $\Pi \rightarrow 0$, the asymptotic solution is,
 \begin{equation}
     \phi(t) \sim \frac{1}{q(\tau)^{1/4}}\, \left(A\, \sin \theta + B\, \cos \theta \right)~, \label{A4}
 \end{equation}
where $\theta = \frac{1}{\Pi}\, \int^{t} \sqrt{q(\tau)}\,{\rm d}\tau $ and $A, B$ are constants to be determined using initial/boundary conditions. The solution can be sinusoidal or exponential type depending on the nature of the `potential' $q(t)$. By substituting respective $q(t)$ terms from Eqs.~(\ref{A1}) and (\ref{A2}) in Eq.~(\ref{A3}) and rearranging, $x'(t)$ and $y'(t)$ can be obtained as in Eqs.~(\ref{eq25}) and (\ref{eq26}) respectively. 

The asymptotic expression Eq.~(\ref{A3}) is valid only away from the `turning point' ($t_{TP}$) at which $q(t_{TP}) = 0$. Thus, from Eq.~(\ref{A2}), $\left\{\frac{\Pi}{t_1}+\left(\Pi-\frac{1}{2}\right)\, \frac{1}{2\, t_1^2} \right\} = 0$ has a solution at $t_1 = (1-2\, \Pi)/(4\, \Pi)$, indicates that there exists a turning point time $t_{TP} = (1-2\, \Pi-4\, St_1)/(4\, \Pi)$ near which the oscillatory solution Eq.~(\ref{eq26}) is not valid (remember $t_1 = t + St_1/\Pi$). Neverthless, away from this turning point time, Eq.~(\ref{eq26}) will be a good approximation. Thus $y'(t_{esc}) \sim 0$ can be asymptotically solved to get escape time estimate when $\Pi \rightarrow 0$ for a SP with $n+m$ is even.
\section{\label{appB}Escape time for condensing droplets estimated using WKB when $\Pi \rightarrow 0$} 
To calculate escape time, here we solve $y'(t_{esc}) \sim 0$.
\subsection{$St_1 > (1-2\, \Pi)/4$}
In this situation, $1-2\, \Pi-4\, St_1 < 0 $ indicates that $t_{TP} < 0$, i.e. the turning point does not exists in positive time, thus the solution Eq.~(\ref{eq26}) is valid in all $t > 0 $. $\chi_{-}$ can be evaluated by performing the integral Eq.~(\ref{eq27}) as $\chi_{-} = F(t+|t_{TP}|)-F(|t_{TP}|)$, where $|t_{TP}| = (-1+2\, \Pi+4\, St_1)/(4\, \Pi) >0 $ and 
\begin{eqnarray}
            F(\tau) = \frac{\sqrt{1-2\, \Pi}}{\Pi}\, \left\{\frac{2\, \sqrt{\Pi\, \tau}}{\sqrt{1-2\, \Pi}}-\tan^{-1}\left[ \frac{2\, \sqrt{\Pi\, \tau}}{\sqrt{1-2\, \Pi}}\right] \right\}~. \nonumber\\
            \label{B1}
\end{eqnarray}
For the initial position $y_0'$ and initial velocity $v_{y0}'$, the constants $C_{11}$ and $C_{12}$ can be evaluated as,
\begin{eqnarray}
      C_{11} = \frac{C_{12}} {G^{+}(|t_{TP}|)}~, \label{B2}\\
      C_{12} = y_0'\, \left(\frac{St_1}{\Pi}\right)^{\gamma/2}\, \Pi^{1/4}\, |t_{TP}|^{1/4}~, \label{B3}
\end{eqnarray}
where (remember $\gamma = -1+\Pi^{-1}$)
\begin{equation}
           G^{\pm}(\tau) = \frac{4\, \sqrt{\Pi\, \tau}}{4\, St_1\, \frac{v_{y0}'}{y_0'} \pm \frac{St_1}{\tau}+2\, (1-\Pi)}~.
           \label{B4}
\end{equation}
From Eq.~(\ref{eq26}), $y'(t_{esc}) = 0$ thus leads to $C_{11}\, \sin \chi_{-} + C_{12}\, \cos \chi_{-} = 0$, can be simplified to 
\begin{eqnarray}
      \tan \left\{ F(t_{esc}+|t_{\text{TP}}|)-F(|t_{\text{TP}}|)\right\} + G^{+}(|t_{\text{TP}}|) = 0~.
      \label{B5}
\end{eqnarray}
The solution $t_{esc}$ of this transcendental expression gives the asymptotic estimate for escape time when $\Pi \rightarrow 0 $ and $St_1 > (1-2\, \Pi)/4$.
\subsection{$St_1 < (1-2\, \Pi)/4$}
In this situation, $1-2\, \Pi-4\, St_1 > 0 $ indicates that $t_{TP}  = (1-2\, \Pi-4\, St_1)/(4\, \Pi) > 0$, i.e. there exists a turning point in positive time, near by which the oscillatory solution Eq.~(\ref{eq26}) is not valid. However, far in time from the turning point (i.e. $t >> t_{TP}$ or $t << t_{TP}$), the solution Eq.~(\ref{eq26}) will be valid. From Eq.~(\ref{eq27}), we can see that the numerical value of $\chi_{-} = i\, (Fh(t_{TP}-t)-Fh(t_{TP}))$ will be purely imaginary for $t \in (0,t_{TP})$ and complex number $\chi_{-} = F(t-t_{TP})-i\,Fh(t_{TP})$ for $t > t_{TP}$, where
\begin{equation}
           Fh(\tau) = \frac{\sqrt{1-2\, \Pi}}{\Pi}\, \left\{\frac{2\, \sqrt{\Pi\, \tau}}{\sqrt{1-2\, \Pi}}-\tanh^{-1}\left[ \frac{2\, \sqrt{\Pi\, \tau}}{\sqrt{1-2\, \Pi}}\right] \right\}~.
           \label{B6}
\end{equation}
Thus, the solution Eq.~(\ref{eq26}) will behave exponentially for $t < t_{TP}$ and can have oscillations only when $t > t_{TP}$. Thus, we conclude that the condensing droplets can hence cross the separatrix by oscillation only when $t > t_{TP}$. (Note that when $\Pi = 0$, this reduces to the case $St < 1/4$ and the corresponding $t_{TP} \rightarrow \infty$, indicates that the particle will never cross the separatrix.)

Using initial position and initial velocity, the constants $C_{11}$ and $C_{12}$ can be evaluated for Eq.~(\ref{eq26}) as,
\begin{eqnarray}
      C_{11} = \frac{-i\, C_{12}} {G^{-}(t_{TP})}~, \label{B7}\\
      C_{12} = \frac{(1+i)}{\sqrt{2}}\, y_0'\, \left(\frac{St_1}{\Pi}\right)^{\gamma/2}\, \Pi^{1/4}\, t_{TP}^{1/4}~. \label{B8}
\end{eqnarray}
However these constants along with Eq.~(\ref{eq26}) will be asymptotically valid estimate of $y'(t)$ only when $t \ll t_{TP}$, and can not be extrapolated for $t \gg t_{TP}$. The constants $C_{11}$ and $C_{12}$ need to be determined separately for this region using appropriate solution matching techniques at $t=t_{TP}$. However, we observed that the real part of the solution Eq.~(\ref{eq26}) along with constants Eq.~(\ref{B7}) and (\ref{B8}) have oscillatory nature and its zeros matches with the zeros of actual asymptote Eq.~(\ref{eq26}) with appropriate constants. Thus simply $\rm{Re}(y'(t_{esc})) \sim 0 $ solved using Eq.~(\ref{eq26}) along with constants Eq.~(\ref{B7}) and (\ref{B8}) can give estimate of escape time when $\Pi \rightarrow 0 $ and $St_1 < (1-2\, \Pi)/4$ as the solution of following transcendental equation,
\begin{eqnarray}
      \tan \left\{ F(t_{esc}-t_{\text{TP}}) \right\}= \nonumber \\ 
      \tanh \left\{Fh(t_{\text{TP}})- 
      \tanh^{-1} G^{-}(t_{\text{TP}}) \right\}~. \label{B9}
\end{eqnarray}
By expanding terms in series for $\Pi \ll 1$, the leading order approximate solution can be obtained as $t_{esc} \sim t_{\text{TP}} + (9\, \pi/ (32\, \sqrt{\Pi}))^{2/3} $.

\nocite{*}

\bibliography{main}

\providecommand{\noopsort}[1]{}\providecommand{\singleletter}[1]{#1}%
\begin{thebibliography}{79}%
\makeatletter
\providecommand \@ifxundefined [1]{%
 \@ifx{#1\undefined}
}%
\providecommand \@ifnum [1]{%
 \ifnum #1\expandafter \@firstoftwo
 \else \expandafter \@secondoftwo
 \fi
}%
\providecommand \@ifx [1]{%
 \ifx #1\expandafter \@firstoftwo
 \else \expandafter \@secondoftwo
 \fi
}%
\providecommand \natexlab [1]{#1}%
\providecommand \enquote  [1]{``#1''}%
\providecommand \bibnamefont  [1]{#1}%
\providecommand \bibfnamefont [1]{#1}%
\providecommand \citenamefont [1]{#1}%
\providecommand \href@noop [0]{\@secondoftwo}%
\providecommand \href [0]{\begingroup \@sanitize@url \@href}%
\providecommand \@href[1]{\@@startlink{#1}\@@href}%
\providecommand \@@href[1]{\endgroup#1\@@endlink}%
\providecommand \@sanitize@url [0]{\catcode `\\12\catcode `\$12\catcode
  `\&12\catcode `\#12\catcode `\^12\catcode `\_12\catcode `\%12\relax}%
\providecommand \@@startlink[1]{}%
\providecommand \@@endlink[0]{}%
\providecommand \url  [0]{\begingroup\@sanitize@url \@url }%
\providecommand \@url [1]{\endgroup\@href {#1}{\urlprefix }}%
\providecommand \urlprefix  [0]{URL }%
\providecommand \Eprint [0]{\href }%
\providecommand \doibase [0]{https://doi.org/}%
\providecommand \selectlanguage [0]{\@gobble}%
\providecommand \bibinfo  [0]{\@secondoftwo}%
\providecommand \bibfield  [0]{\@secondoftwo}%
\providecommand \translation [1]{[#1]}%
\providecommand \BibitemOpen [0]{}%
\providecommand \bibitemStop [0]{}%
\providecommand \bibitemNoStop [0]{.\EOS\space}%
\providecommand \EOS [0]{\spacefactor3000\relax}%
\providecommand \BibitemShut  [1]{\csname bibitem#1\endcsname}%
\let\auto@bib@innerbib\@empty
\bibitem [{\citenamefont {Clift}\ \emph {et~al.}(2005)\citenamefont {Clift},
  \citenamefont {Grace},\ and\ \citenamefont {Weber}}]{clift_grace}%
  \BibitemOpen
  \bibfield  {author} {\bibinfo {author} {\bibfnamefont {R.}~\bibnamefont
  {Clift}}, \bibinfo {author} {\bibfnamefont {J.~R.}\ \bibnamefont {Grace}},\
  and\ \bibinfo {author} {\bibfnamefont {M.~E.}\ \bibnamefont {Weber}},\
  }\href@noop {} {\emph {\bibinfo {title} {Bubbles, drops, and particles}}}\
  (\bibinfo  {publisher} {Courier Corporation},\ \bibinfo {year}
  {2005})\BibitemShut {NoStop}%
\bibitem [{\citenamefont {Maxey}(1987)}]{maxey1987motion}%
  \BibitemOpen
  \bibfield  {author} {\bibinfo {author} {\bibfnamefont {M.~R.}\ \bibnamefont
  {Maxey}},\ }\bibfield  {title} {\bibinfo {title} {The motion of small
  spherical particles in a cellular flow field},\ }\href@noop {} {\bibfield
  {journal} {\bibinfo  {journal} {The Physics of fluids}\ }\textbf {\bibinfo
  {volume} {30}},\ \bibinfo {pages} {1915} (\bibinfo {year}
  {1987})}\BibitemShut {NoStop}%
\bibitem [{\citenamefont {Vincent}\ and\ \citenamefont
  {Meneguzzi}(1994{\natexlab{a}})}]{vincent_meneguzzi_1994}%
  \BibitemOpen
  \bibfield  {author} {\bibinfo {author} {\bibfnamefont {A.}~\bibnamefont
  {Vincent}}\ and\ \bibinfo {author} {\bibfnamefont {M.}~\bibnamefont
  {Meneguzzi}},\ }\bibfield  {title} {\bibinfo {title} {The dynamics of
  vorticity tubes in homogeneous turbulence},\ }\href
  {https://doi.org/10.1017/S0022112094003319} {\bibfield  {journal} {\bibinfo
  {journal} {Journal of Fluid Mechanics}\ }\textbf {\bibinfo {volume} {258}},\
  \bibinfo {pages} {245–254} (\bibinfo {year}
  {1994}{\natexlab{a}})}\BibitemShut {NoStop}%
\bibitem [{\citenamefont {Bec}(2003)}]{bec2003fractal}%
  \BibitemOpen
  \bibfield  {author} {\bibinfo {author} {\bibfnamefont {J.}~\bibnamefont
  {Bec}},\ }\bibfield  {title} {\bibinfo {title} {Fractal clustering of
  inertial particles in random flows},\ }\href@noop {} {\bibfield  {journal}
  {\bibinfo  {journal} {Physics of fluids}\ }\textbf {\bibinfo {volume} {15}},\
  \bibinfo {pages} {L81} (\bibinfo {year} {2003})}\BibitemShut {NoStop}%
\bibitem [{\citenamefont {Chen}\ \emph {et~al.}(2006)\citenamefont {Chen},
  \citenamefont {Goto},\ and\ \citenamefont {Vassilicos}}]{chen2006turbulent}%
  \BibitemOpen
  \bibfield  {author} {\bibinfo {author} {\bibfnamefont {L.}~\bibnamefont
  {Chen}}, \bibinfo {author} {\bibfnamefont {S.}~\bibnamefont {Goto}},\ and\
  \bibinfo {author} {\bibfnamefont {J.~C.}\ \bibnamefont {Vassilicos}},\
  }\bibfield  {title} {\bibinfo {title} {Turbulent clustering of stagnation
  points and inertial particles},\ }\href@noop {} {\bibfield  {journal}
  {\bibinfo  {journal} {Journal of Fluid Mechanics}\ }\textbf {\bibinfo
  {volume} {553}},\ \bibinfo {pages} {143} (\bibinfo {year}
  {2006})}\BibitemShut {NoStop}%
\bibitem [{\citenamefont {Angilella}(2010)}]{angilella2010dust}%
  \BibitemOpen
  \bibfield  {author} {\bibinfo {author} {\bibfnamefont {J.-R.}\ \bibnamefont
  {Angilella}},\ }\bibfield  {title} {\bibinfo {title} {Dust trapping in vortex
  pairs},\ }\href@noop {} {\bibfield  {journal} {\bibinfo  {journal} {Physica.
  D}\ }\textbf {\bibinfo {volume} {239}},\ \bibinfo {pages} {1789} (\bibinfo
  {year} {2010})}\BibitemShut {NoStop}%
\bibitem [{\citenamefont {Monchaux}\ \emph {et~al.}(2012)\citenamefont
  {Monchaux}, \citenamefont {Bourgoin},\ and\ \citenamefont
  {Cartellier}}]{monchaux2012analyzing}%
  \BibitemOpen
  \bibfield  {author} {\bibinfo {author} {\bibfnamefont {R.}~\bibnamefont
  {Monchaux}}, \bibinfo {author} {\bibfnamefont {M.}~\bibnamefont {Bourgoin}},\
  and\ \bibinfo {author} {\bibfnamefont {A.}~\bibnamefont {Cartellier}},\
  }\bibfield  {title} {\bibinfo {title} {Analyzing preferential concentration
  and clustering of inertial particles in turbulence},\ }\href@noop {}
  {\bibfield  {journal} {\bibinfo  {journal} {International Journal of
  Multiphase Flow}\ }\textbf {\bibinfo {volume} {40}},\ \bibinfo {pages} {1}
  (\bibinfo {year} {2012})}\BibitemShut {NoStop}%
\bibitem [{\citenamefont {Petersen}\ \emph {et~al.}(2019)\citenamefont
  {Petersen}, \citenamefont {Baker},\ and\ \citenamefont
  {Coletti}}]{petersen2019experimental}%
  \BibitemOpen
  \bibfield  {author} {\bibinfo {author} {\bibfnamefont {A.~J.}\ \bibnamefont
  {Petersen}}, \bibinfo {author} {\bibfnamefont {L.}~\bibnamefont {Baker}},\
  and\ \bibinfo {author} {\bibfnamefont {F.}~\bibnamefont {Coletti}},\
  }\bibfield  {title} {\bibinfo {title} {Experimental study of inertial
  particles clustering and settling in homogeneous turbulence},\ }\href@noop {}
  {\bibfield  {journal} {\bibinfo  {journal} {Journal of Fluid Mechanics}\
  }\textbf {\bibinfo {volume} {864}},\ \bibinfo {pages} {925} (\bibinfo {year}
  {2019})}\BibitemShut {NoStop}%
\bibitem [{\citenamefont {Yoshimoto}\ and\ \citenamefont
  {Goto}(2007)}]{yoshimoto2007self}%
  \BibitemOpen
  \bibfield  {author} {\bibinfo {author} {\bibfnamefont {H.}~\bibnamefont
  {Yoshimoto}}\ and\ \bibinfo {author} {\bibfnamefont {S.}~\bibnamefont
  {Goto}},\ }\bibfield  {title} {\bibinfo {title} {Self-similar clustering of
  inertial particles in homogeneous turbulence},\ }\href@noop {} {\bibfield
  {journal} {\bibinfo  {journal} {Journal of Fluid Mechanics}\ }\textbf
  {\bibinfo {volume} {577}},\ \bibinfo {pages} {275} (\bibinfo {year}
  {2007})}\BibitemShut {NoStop}%
\bibitem [{\citenamefont {Bragg}\ and\ \citenamefont
  {Collins}(2014)}]{bragg2014new}%
  \BibitemOpen
  \bibfield  {author} {\bibinfo {author} {\bibfnamefont {A.~D.}\ \bibnamefont
  {Bragg}}\ and\ \bibinfo {author} {\bibfnamefont {L.~R.}\ \bibnamefont
  {Collins}},\ }\bibfield  {title} {\bibinfo {title} {New insights from
  comparing statistical theories for inertial particles in turbulence: I.
  spatial distribution of particles},\ }\href@noop {} {\bibfield  {journal}
  {\bibinfo  {journal} {New Journal of Physics}\ }\textbf {\bibinfo {volume}
  {16}},\ \bibinfo {pages} {055013} (\bibinfo {year} {2014})}\BibitemShut
  {NoStop}%
\bibitem [{\citenamefont {Goto}\ and\ \citenamefont
  {Vassilicos}(2006)}]{goto2006self}%
  \BibitemOpen
  \bibfield  {author} {\bibinfo {author} {\bibfnamefont {S.}~\bibnamefont
  {Goto}}\ and\ \bibinfo {author} {\bibfnamefont {J.~C.}\ \bibnamefont
  {Vassilicos}},\ }\bibfield  {title} {\bibinfo {title} {Self-similar
  clustering of inertial particles and zero-acceleration points in fully
  developed two-dimensional turbulence},\ }\href@noop {} {\bibfield  {journal}
  {\bibinfo  {journal} {Physics of Fluids}\ }\textbf {\bibinfo {volume} {18}},\
  \bibinfo {pages} {115103} (\bibinfo {year} {2006})}\BibitemShut {NoStop}%
\bibitem [{\citenamefont {Saw}\ \emph {et~al.}(2012)\citenamefont {Saw},
  \citenamefont {Shaw}, \citenamefont {Salazar},\ and\ \citenamefont
  {Collins}}]{saw2012spatial}%
  \BibitemOpen
  \bibfield  {author} {\bibinfo {author} {\bibfnamefont {E.-W.}\ \bibnamefont
  {Saw}}, \bibinfo {author} {\bibfnamefont {R.~A.}\ \bibnamefont {Shaw}},
  \bibinfo {author} {\bibfnamefont {J.~P.}\ \bibnamefont {Salazar}},\ and\
  \bibinfo {author} {\bibfnamefont {L.~R.}\ \bibnamefont {Collins}},\
  }\bibfield  {title} {\bibinfo {title} {Spatial clustering of polydisperse
  inertial particles in turbulence: ii. comparing simulation with experiment},\
  }\href@noop {} {\bibfield  {journal} {\bibinfo  {journal} {New Journal of
  Physics}\ }\textbf {\bibinfo {volume} {14}},\ \bibinfo {pages} {105031}
  (\bibinfo {year} {2012})}\BibitemShut {NoStop}%
\bibitem [{\citenamefont {Baker}\ \emph {et~al.}(2017)\citenamefont {Baker},
  \citenamefont {Frankel}, \citenamefont {Mani},\ and\ \citenamefont
  {Coletti}}]{baker2017coherent}%
  \BibitemOpen
  \bibfield  {author} {\bibinfo {author} {\bibfnamefont {L.}~\bibnamefont
  {Baker}}, \bibinfo {author} {\bibfnamefont {A.}~\bibnamefont {Frankel}},
  \bibinfo {author} {\bibfnamefont {A.}~\bibnamefont {Mani}},\ and\ \bibinfo
  {author} {\bibfnamefont {F.}~\bibnamefont {Coletti}},\ }\bibfield  {title}
  {\bibinfo {title} {Coherent clusters of inertial particles in homogeneous
  turbulence},\ }\href@noop {} {\bibfield  {journal} {\bibinfo  {journal}
  {Journal of Fluid Mechanics}\ }\textbf {\bibinfo {volume} {833}},\ \bibinfo
  {pages} {364} (\bibinfo {year} {2017})}\BibitemShut {NoStop}%
\bibitem [{\citenamefont {Ravichandran}\ \emph {et~al.}(2014)\citenamefont
  {Ravichandran}, \citenamefont {Perlekar},\ and\ \citenamefont
  {Govindarajan}}]{ravichandran2014attracting}%
  \BibitemOpen
  \bibfield  {author} {\bibinfo {author} {\bibfnamefont {S.}~\bibnamefont
  {Ravichandran}}, \bibinfo {author} {\bibfnamefont {P.}~\bibnamefont
  {Perlekar}},\ and\ \bibinfo {author} {\bibfnamefont {R.}~\bibnamefont
  {Govindarajan}},\ }\bibfield  {title} {\bibinfo {title} {Attracting fixed
  points for heavy particles in the vicinity of a vortex pair},\ }\href@noop {}
  {\bibfield  {journal} {\bibinfo  {journal} {Physics of Fluids}\ }\textbf
  {\bibinfo {volume} {26}},\ \bibinfo {pages} {013303} (\bibinfo {year}
  {2014})}\BibitemShut {NoStop}%
\bibitem [{\citenamefont {Gustavsson}\ and\ \citenamefont
  {Mehlig}(2016{\natexlab{a}})}]{gustavsson2016statistical}%
  \BibitemOpen
  \bibfield  {author} {\bibinfo {author} {\bibfnamefont {K.}~\bibnamefont
  {Gustavsson}}\ and\ \bibinfo {author} {\bibfnamefont {B.}~\bibnamefont
  {Mehlig}},\ }\bibfield  {title} {\bibinfo {title} {Statistical models for
  spatial patterns of heavy particles in turbulence},\ }\href@noop {}
  {\bibfield  {journal} {\bibinfo  {journal} {Advances in Physics}\ }\textbf
  {\bibinfo {volume} {65}},\ \bibinfo {pages} {1} (\bibinfo {year}
  {2016}{\natexlab{a}})}\BibitemShut {NoStop}%
\bibitem [{\citenamefont {Wilkinson}\ and\ \citenamefont
  {Mehlig}(2005)}]{wilkinson2005caustics}%
  \BibitemOpen
  \bibfield  {author} {\bibinfo {author} {\bibfnamefont {M.}~\bibnamefont
  {Wilkinson}}\ and\ \bibinfo {author} {\bibfnamefont {B.}~\bibnamefont
  {Mehlig}},\ }\bibfield  {title} {\bibinfo {title} {Caustics in turbulent
  aerosols},\ }\href@noop {} {\bibfield  {journal} {\bibinfo  {journal} {EPL
  (Europhysics Letters)}\ }\textbf {\bibinfo {volume} {71}},\ \bibinfo {pages}
  {186} (\bibinfo {year} {2005})}\BibitemShut {NoStop}%
\bibitem [{\citenamefont {Gustavsson}\ \emph {et~al.}(2012)\citenamefont
  {Gustavsson}, \citenamefont {Meneguz}, \citenamefont {Reeks},\ and\
  \citenamefont {Mehlig}}]{gustavsson2012inertial}%
  \BibitemOpen
  \bibfield  {author} {\bibinfo {author} {\bibfnamefont {K.}~\bibnamefont
  {Gustavsson}}, \bibinfo {author} {\bibfnamefont {E.}~\bibnamefont {Meneguz}},
  \bibinfo {author} {\bibfnamefont {M.}~\bibnamefont {Reeks}},\ and\ \bibinfo
  {author} {\bibfnamefont {B.}~\bibnamefont {Mehlig}},\ }\bibfield  {title}
  {\bibinfo {title} {Inertial-particle dynamics in turbulent flows: caustics,
  concentration fluctuations and random uncorrelated motion},\ }\href@noop {}
  {\bibfield  {journal} {\bibinfo  {journal} {New Journal of Physics}\ }\textbf
  {\bibinfo {volume} {14}},\ \bibinfo {pages} {115017} (\bibinfo {year}
  {2012})}\BibitemShut {NoStop}%
\bibitem [{\citenamefont {Bec}\ \emph {et~al.}(2010)\citenamefont {Bec},
  \citenamefont {Biferale}, \citenamefont {Cencini}, \citenamefont {Lanotte},\
  and\ \citenamefont {Toschi}}]{bec2010intermittency}%
  \BibitemOpen
  \bibfield  {author} {\bibinfo {author} {\bibfnamefont {J.}~\bibnamefont
  {Bec}}, \bibinfo {author} {\bibfnamefont {L.}~\bibnamefont {Biferale}},
  \bibinfo {author} {\bibfnamefont {M.}~\bibnamefont {Cencini}}, \bibinfo
  {author} {\bibfnamefont {A.~S.}\ \bibnamefont {Lanotte}},\ and\ \bibinfo
  {author} {\bibfnamefont {F.}~\bibnamefont {Toschi}},\ }\bibfield  {title}
  {\bibinfo {title} {Intermittency in the velocity distribution of heavy
  particles in turbulence},\ }\href@noop {} {\bibfield  {journal} {\bibinfo
  {journal} {Journal of Fluid Mechanics}\ }\textbf {\bibinfo {volume} {646}},\
  \bibinfo {pages} {527} (\bibinfo {year} {2010})}\BibitemShut {NoStop}%
\bibitem [{\citenamefont {Gustavsson}\ \emph {et~al.}(2014)\citenamefont
  {Gustavsson}, \citenamefont {Einarsson},\ and\ \citenamefont
  {Mehlig}}]{gustavsson2014tumbling}%
  \BibitemOpen
  \bibfield  {author} {\bibinfo {author} {\bibfnamefont {K.}~\bibnamefont
  {Gustavsson}}, \bibinfo {author} {\bibfnamefont {J.}~\bibnamefont
  {Einarsson}},\ and\ \bibinfo {author} {\bibfnamefont {B.}~\bibnamefont
  {Mehlig}},\ }\bibfield  {title} {\bibinfo {title} {Tumbling of small
  axisymmetric particles in random and turbulent flows},\ }\href@noop {}
  {\bibfield  {journal} {\bibinfo  {journal} {Physical review letters}\
  }\textbf {\bibinfo {volume} {112}},\ \bibinfo {pages} {014501} (\bibinfo
  {year} {2014})}\BibitemShut {NoStop}%
\bibitem [{\citenamefont {Ravichandran}\ and\ \citenamefont
  {Govindarajan}(2015)}]{ravichandran2015caustics}%
  \BibitemOpen
  \bibfield  {author} {\bibinfo {author} {\bibfnamefont {S.}~\bibnamefont
  {Ravichandran}}\ and\ \bibinfo {author} {\bibfnamefont {R.}~\bibnamefont
  {Govindarajan}},\ }\bibfield  {title} {\bibinfo {title} {Caustics and
  clustering in the vicinity of a vortex},\ }\href@noop {} {\bibfield
  {journal} {\bibinfo  {journal} {Physics of Fluids}\ }\textbf {\bibinfo
  {volume} {27}},\ \bibinfo {pages} {033305} (\bibinfo {year}
  {2015})}\BibitemShut {NoStop}%
\bibitem [{\citenamefont {Gustavsson}\ and\ \citenamefont
  {Mehlig}(2016{\natexlab{b}})}]{gustavsson2016statistical1}%
  \BibitemOpen
  \bibfield  {author} {\bibinfo {author} {\bibfnamefont {K.}~\bibnamefont
  {Gustavsson}}\ and\ \bibinfo {author} {\bibfnamefont {B.}~\bibnamefont
  {Mehlig}},\ }\bibfield  {title} {\bibinfo {title} {Statistical model for
  collisions and recollisions of inertial particles in mixing flows},\
  }\href@noop {} {\bibfield  {journal} {\bibinfo  {journal} {The European
  Physical Journal E}\ }\textbf {\bibinfo {volume} {39}},\ \bibinfo {pages} {1}
  (\bibinfo {year} {2016}{\natexlab{b}})}\BibitemShut {NoStop}%
\bibitem [{\citenamefont {Deepu}\ \emph {et~al.}(2017)\citenamefont {Deepu},
  \citenamefont {Ravichandran},\ and\ \citenamefont
  {Govindarajan}}]{deepu2017caustics}%
  \BibitemOpen
  \bibfield  {author} {\bibinfo {author} {\bibfnamefont {P.}~\bibnamefont
  {Deepu}}, \bibinfo {author} {\bibfnamefont {S.}~\bibnamefont
  {Ravichandran}},\ and\ \bibinfo {author} {\bibfnamefont {R.}~\bibnamefont
  {Govindarajan}},\ }\bibfield  {title} {\bibinfo {title} {Caustics-induced
  coalescence of small droplets near a vortex},\ }\href@noop {} {\bibfield
  {journal} {\bibinfo  {journal} {Physical Review Fluids}\ }\textbf {\bibinfo
  {volume} {2}},\ \bibinfo {pages} {024305} (\bibinfo {year}
  {2017})}\BibitemShut {NoStop}%
\bibitem [{\citenamefont {Elghobashi}(1994)}]{elghobashi1994predicting}%
  \BibitemOpen
  \bibfield  {author} {\bibinfo {author} {\bibfnamefont {S.}~\bibnamefont
  {Elghobashi}},\ }\bibfield  {title} {\bibinfo {title} {On predicting
  particle-laden turbulent flows},\ }\href@noop {} {\bibfield  {journal}
  {\bibinfo  {journal} {Applied scientific research}\ }\textbf {\bibinfo
  {volume} {52}},\ \bibinfo {pages} {309} (\bibinfo {year} {1994})}\BibitemShut
  {NoStop}%
\bibitem [{\citenamefont {Candelier}\ \emph {et~al.}(2016)\citenamefont
  {Candelier}, \citenamefont {Einarsson},\ and\ \citenamefont
  {Mehlig}}]{candelier2016angular}%
  \BibitemOpen
  \bibfield  {author} {\bibinfo {author} {\bibfnamefont {F.}~\bibnamefont
  {Candelier}}, \bibinfo {author} {\bibfnamefont {J.}~\bibnamefont
  {Einarsson}},\ and\ \bibinfo {author} {\bibfnamefont {B.}~\bibnamefont
  {Mehlig}},\ }\bibfield  {title} {\bibinfo {title} {Angular dynamics of a
  small particle in turbulence},\ }\href@noop {} {\bibfield  {journal}
  {\bibinfo  {journal} {Physical review letters}\ }\textbf {\bibinfo {volume}
  {117}},\ \bibinfo {pages} {204501} (\bibinfo {year} {2016})}\BibitemShut
  {NoStop}%
\bibitem [{\citenamefont {Richter}\ and\ \citenamefont
  {Sullivan}(2014)}]{richter2014modification}%
  \BibitemOpen
  \bibfield  {author} {\bibinfo {author} {\bibfnamefont {D.~H.}\ \bibnamefont
  {Richter}}\ and\ \bibinfo {author} {\bibfnamefont {P.~P.}\ \bibnamefont
  {Sullivan}},\ }\bibfield  {title} {\bibinfo {title} {Modification of
  near-wall coherent structures by inertial particles},\ }\href@noop {}
  {\bibfield  {journal} {\bibinfo  {journal} {Physics of Fluids}\ }\textbf
  {\bibinfo {volume} {26}},\ \bibinfo {pages} {103304} (\bibinfo {year}
  {2014})}\BibitemShut {NoStop}%
\bibitem [{\citenamefont {Muramulla}\ \emph {et~al.}(2020)\citenamefont
  {Muramulla}, \citenamefont {Tyagi}, \citenamefont {Goswami},\ and\
  \citenamefont {Kumaran}}]{muramulla2020disruption}%
  \BibitemOpen
  \bibfield  {author} {\bibinfo {author} {\bibfnamefont {P.}~\bibnamefont
  {Muramulla}}, \bibinfo {author} {\bibfnamefont {A.}~\bibnamefont {Tyagi}},
  \bibinfo {author} {\bibfnamefont {P.~S.}\ \bibnamefont {Goswami}},\ and\
  \bibinfo {author} {\bibfnamefont {V.}~\bibnamefont {Kumaran}},\ }\bibfield
  {title} {\bibinfo {title} {Disruption of turbulence due to particle loading
  in a dilute gas--particle suspension},\ }\href@noop {} {\bibfield  {journal}
  {\bibinfo  {journal} {Journal of Fluid Mechanics}\ }\textbf {\bibinfo
  {volume} {889}} (\bibinfo {year} {2020})}\BibitemShut {NoStop}%
\bibitem [{\citenamefont {Veron}(2015)}]{veron2015ocean}%
  \BibitemOpen
  \bibfield  {author} {\bibinfo {author} {\bibfnamefont {F.}~\bibnamefont
  {Veron}},\ }\bibfield  {title} {\bibinfo {title} {Ocean spray},\ }\href@noop
  {} {\bibfield  {journal} {\bibinfo  {journal} {Annual Review of Fluid
  Mechanics}\ }\textbf {\bibinfo {volume} {47}},\ \bibinfo {pages} {507}
  (\bibinfo {year} {2015})}\BibitemShut {NoStop}%
\bibitem [{\citenamefont {Narasimha}(2012)}]{narasimha2012cumulus}%
  \BibitemOpen
  \bibfield  {author} {\bibinfo {author} {\bibfnamefont {R.}~\bibnamefont
  {Narasimha}},\ }\bibfield  {title} {\bibinfo {title} {Cumulus clouds and
  convective boundary layers: a tropical perspective on two turbulent shear
  flows},\ }\href@noop {} {\bibfield  {journal} {\bibinfo  {journal} {Journal
  of Turbulence}\ ,\ \bibinfo {pages} {N47}} (\bibinfo {year}
  {2012})}\BibitemShut {NoStop}%
\bibitem [{\citenamefont {Chong}\ \emph {et~al.}(2021)\citenamefont {Chong},
  \citenamefont {Ng}, \citenamefont {Hori}, \citenamefont {Yang}, \citenamefont
  {Verzicco},\ and\ \citenamefont {Lohse}}]{lohse2021}%
  \BibitemOpen
  \bibfield  {author} {\bibinfo {author} {\bibfnamefont {K.~L.}\ \bibnamefont
  {Chong}}, \bibinfo {author} {\bibfnamefont {C.~S.}\ \bibnamefont {Ng}},
  \bibinfo {author} {\bibfnamefont {N.}~\bibnamefont {Hori}}, \bibinfo {author}
  {\bibfnamefont {R.}~\bibnamefont {Yang}}, \bibinfo {author} {\bibfnamefont
  {R.}~\bibnamefont {Verzicco}},\ and\ \bibinfo {author} {\bibfnamefont
  {D.}~\bibnamefont {Lohse}},\ }\bibfield  {title} {\bibinfo {title} {Extended
  lifetime of respiratory droplets in a turbulent vapor puff and its
  implications on airborne disease transmission},\ }\href@noop {} {\bibfield
  {journal} {\bibinfo  {journal} {Physical review letters}\ }\textbf {\bibinfo
  {volume} {126}},\ \bibinfo {pages} {034502} (\bibinfo {year}
  {2021})}\BibitemShut {NoStop}%
\bibitem [{\citenamefont {Rosti}\ \emph {et~al.}(2021)\citenamefont {Rosti},
  \citenamefont {Cavaiola}, \citenamefont {Olivieri}, \citenamefont
  {Seminara},\ and\ \citenamefont {Mazzino}}]{rosti2021turbulence}%
  \BibitemOpen
  \bibfield  {author} {\bibinfo {author} {\bibfnamefont {M.~E.}\ \bibnamefont
  {Rosti}}, \bibinfo {author} {\bibfnamefont {M.}~\bibnamefont {Cavaiola}},
  \bibinfo {author} {\bibfnamefont {S.}~\bibnamefont {Olivieri}}, \bibinfo
  {author} {\bibfnamefont {A.}~\bibnamefont {Seminara}},\ and\ \bibinfo
  {author} {\bibfnamefont {A.}~\bibnamefont {Mazzino}},\ }\bibfield  {title}
  {\bibinfo {title} {Turbulence role in the fate of virus-containing droplets
  in violent expiratory events},\ }\href@noop {} {\bibfield  {journal}
  {\bibinfo  {journal} {Physical Review Research}\ }\textbf {\bibinfo {volume}
  {3}},\ \bibinfo {pages} {013091} (\bibinfo {year} {2021})}\BibitemShut
  {NoStop}%
\bibitem [{\citenamefont {Diwan}\ \emph {et~al.}(2020)\citenamefont {Diwan},
  \citenamefont {Ravichandran}, \citenamefont {Govindarajan},\ and\
  \citenamefont {Narasimha}}]{diwan2020}%
  \BibitemOpen
  \bibfield  {author} {\bibinfo {author} {\bibfnamefont {S.~S.}\ \bibnamefont
  {Diwan}}, \bibinfo {author} {\bibfnamefont {S.}~\bibnamefont {Ravichandran}},
  \bibinfo {author} {\bibfnamefont {R.}~\bibnamefont {Govindarajan}},\ and\
  \bibinfo {author} {\bibfnamefont {R.}~\bibnamefont {Narasimha}},\ }\bibfield
  {title} {\bibinfo {title} {Understanding transmission dynamics of
  covid-19-type infections by direct numerical simulations of cough/sneeze
  flows},\ }\href@noop {} {\bibfield  {journal} {\bibinfo  {journal}
  {Transactions of the Indian National Academy of Engineering}\ }\textbf
  {\bibinfo {volume} {5}},\ \bibinfo {pages} {255} (\bibinfo {year}
  {2020})}\BibitemShut {NoStop}%
\bibitem [{\citenamefont {Singhal}\ \emph {et~al.}(2021)\citenamefont
  {Singhal}, \citenamefont {Ravichandran}, \citenamefont {Govindarajan},\ and\
  \citenamefont {Diwan}}]{singhal2021}%
  \BibitemOpen
  \bibfield  {author} {\bibinfo {author} {\bibfnamefont {R.}~\bibnamefont
  {Singhal}}, \bibinfo {author} {\bibfnamefont {S.}~\bibnamefont
  {Ravichandran}}, \bibinfo {author} {\bibfnamefont {R.}~\bibnamefont
  {Govindarajan}},\ and\ \bibinfo {author} {\bibfnamefont {S.~S.}\ \bibnamefont
  {Diwan}},\ }\bibfield  {title} {\bibinfo {title} {Virus transmission by
  aerosol transport during short conversations},\ }\href@noop {} {\bibfield
  {journal} {\bibinfo  {journal} {arXiv preprint arXiv:2103.16415}\ } (\bibinfo
  {year} {2021})}\BibitemShut {NoStop}%
\bibitem [{\citenamefont {Pruppacher}\ and\ \citenamefont
  {Klett}(2010)}]{pruppacher2010microstructure}%
  \BibitemOpen
  \bibfield  {author} {\bibinfo {author} {\bibfnamefont {H.~R.}\ \bibnamefont
  {Pruppacher}}\ and\ \bibinfo {author} {\bibfnamefont {J.~D.}\ \bibnamefont
  {Klett}},\ }\bibfield  {title} {\bibinfo {title} {Microstructure of
  atmospheric clouds and precipitation},\ }in\ \href@noop {} {\emph {\bibinfo
  {booktitle} {Microphysics of Clouds and Precipitation}}}\ (\bibinfo
  {publisher} {Springer},\ \bibinfo {year} {2010})\ pp.\ \bibinfo {pages}
  {10--73}\BibitemShut {NoStop}%
\bibitem [{\citenamefont {Andreas}(1989)}]{andreas1989thermal}%
  \BibitemOpen
  \bibfield  {author} {\bibinfo {author} {\bibfnamefont {E.~L.}\ \bibnamefont
  {Andreas}},\ }in\ \href@noop {} {\emph {\bibinfo {booktitle} {Thermal and
  size evolution of sea spray droplets}}}\ (\bibinfo  {publisher} {Cold Regions
  Research and Engineering Laboratory (US)},\ \bibinfo {year}
  {1989})\BibitemShut {NoStop}%
\bibitem [{\citenamefont {Helgans}\ and\ \citenamefont
  {Richter}(2016)}]{helgans2016turbulent}%
  \BibitemOpen
  \bibfield  {author} {\bibinfo {author} {\bibfnamefont {B.}~\bibnamefont
  {Helgans}}\ and\ \bibinfo {author} {\bibfnamefont {D.~H.}\ \bibnamefont
  {Richter}},\ }\bibfield  {title} {\bibinfo {title} {Turbulent latent and
  sensible heat flux in the presence of evaporative droplets},\ }\href@noop {}
  {\bibfield  {journal} {\bibinfo  {journal} {International Journal of
  Multiphase Flow}\ }\textbf {\bibinfo {volume} {78}},\ \bibinfo {pages} {1}
  (\bibinfo {year} {2016})}\BibitemShut {NoStop}%
\bibitem [{\citenamefont {Shaw}\ \emph {et~al.}(1998)\citenamefont {Shaw},
  \citenamefont {Reade}, \citenamefont {Collins},\ and\ \citenamefont
  {Verlinde}}]{shaw1998preferential}%
  \BibitemOpen
  \bibfield  {author} {\bibinfo {author} {\bibfnamefont {R.~A.}\ \bibnamefont
  {Shaw}}, \bibinfo {author} {\bibfnamefont {W.~C.}\ \bibnamefont {Reade}},
  \bibinfo {author} {\bibfnamefont {L.~R.}\ \bibnamefont {Collins}},\ and\
  \bibinfo {author} {\bibfnamefont {J.}~\bibnamefont {Verlinde}},\ }\bibfield
  {title} {\bibinfo {title} {Preferential concentration of cloud droplets by
  turbulence: Effects on the early evolution of cumulus cloud droplet
  spectra},\ }\href@noop {} {\bibfield  {journal} {\bibinfo  {journal} {Journal
  of the atmospheric sciences}\ }\textbf {\bibinfo {volume} {55}},\ \bibinfo
  {pages} {1965} (\bibinfo {year} {1998})}\BibitemShut {NoStop}%
\bibitem [{\citenamefont {Vaillancourt}\ \emph {et~al.}(2002)\citenamefont
  {Vaillancourt}, \citenamefont {Yau}, \citenamefont {Bartello},\ and\
  \citenamefont {Grabowski}}]{vaillancourt2002microscopic}%
  \BibitemOpen
  \bibfield  {author} {\bibinfo {author} {\bibfnamefont {P.~A.}\ \bibnamefont
  {Vaillancourt}}, \bibinfo {author} {\bibfnamefont {M.~K.}\ \bibnamefont
  {Yau}}, \bibinfo {author} {\bibfnamefont {P.}~\bibnamefont {Bartello}},\ and\
  \bibinfo {author} {\bibfnamefont {W.~W.}\ \bibnamefont {Grabowski}},\
  }\bibfield  {title} {\bibinfo {title} {Microscopic approach to cloud droplet
  growth by condensation. part ii: Turbulence, clustering, and condensational
  growth},\ }\href@noop {} {\bibfield  {journal} {\bibinfo  {journal} {Journal
  of the atmospheric sciences}\ }\textbf {\bibinfo {volume} {59}},\ \bibinfo
  {pages} {3421} (\bibinfo {year} {2002})}\BibitemShut {NoStop}%
\bibitem [{\citenamefont {Sardina}\ \emph {et~al.}(2015)\citenamefont
  {Sardina}, \citenamefont {Picano}, \citenamefont {Brandt},\ and\
  \citenamefont {Caballero}}]{sardina2015continuous}%
  \BibitemOpen
  \bibfield  {author} {\bibinfo {author} {\bibfnamefont {G.}~\bibnamefont
  {Sardina}}, \bibinfo {author} {\bibfnamefont {F.}~\bibnamefont {Picano}},
  \bibinfo {author} {\bibfnamefont {L.}~\bibnamefont {Brandt}},\ and\ \bibinfo
  {author} {\bibfnamefont {R.}~\bibnamefont {Caballero}},\ }\bibfield  {title}
  {\bibinfo {title} {Continuous growth of droplet size variance due to
  condensation in turbulent clouds},\ }\href@noop {} {\bibfield  {journal}
  {\bibinfo  {journal} {Physical review letters}\ }\textbf {\bibinfo {volume}
  {115}},\ \bibinfo {pages} {184501} (\bibinfo {year} {2015})}\BibitemShut
  {NoStop}%
\bibitem [{\citenamefont {Drossinos}\ and\ \citenamefont
  {Reeks}(2005)}]{drossinos2005brownian}%
  \BibitemOpen
  \bibfield  {author} {\bibinfo {author} {\bibfnamefont {Y.}~\bibnamefont
  {Drossinos}}\ and\ \bibinfo {author} {\bibfnamefont {M.~W.}\ \bibnamefont
  {Reeks}},\ }\bibfield  {title} {\bibinfo {title} {Brownian motion of
  finite-inertia particles in a simple shear flow},\ }\href@noop {} {\bibfield
  {journal} {\bibinfo  {journal} {Physical Review E}\ }\textbf {\bibinfo
  {volume} {71}},\ \bibinfo {pages} {031113} (\bibinfo {year}
  {2005})}\BibitemShut {NoStop}%
\bibitem [{\citenamefont {Renaud}\ and\ \citenamefont
  {Vanneste}(2020)}]{renaud2020dispersion}%
  \BibitemOpen
  \bibfield  {author} {\bibinfo {author} {\bibfnamefont {A.}~\bibnamefont
  {Renaud}}\ and\ \bibinfo {author} {\bibfnamefont {J.}~\bibnamefont
  {Vanneste}},\ }\bibfield  {title} {\bibinfo {title} {Dispersion of inertial
  particles in cellular flows in the small-stokes, large-péclet regime},\
  }\href@noop {} {\bibfield  {journal} {\bibinfo  {journal} {Journal of Fluid
  Mechanics}\ }\textbf {\bibinfo {volume} {903}} (\bibinfo {year}
  {2020})}\BibitemShut {NoStop}%
\bibitem [{\citenamefont {Pavliotis}\ and\ \citenamefont
  {Stuart}(2005)}]{pavliotis2005periodic}%
  \BibitemOpen
  \bibfield  {author} {\bibinfo {author} {\bibfnamefont {G.~A.}\ \bibnamefont
  {Pavliotis}}\ and\ \bibinfo {author} {\bibfnamefont {A.~M.}\ \bibnamefont
  {Stuart}},\ }\bibfield  {title} {\bibinfo {title} {Periodic homogenization
  for inertial particles},\ }\href@noop {} {\bibfield  {journal} {\bibinfo
  {journal} {Physica D: Nonlinear Phenomena}\ }\textbf {\bibinfo {volume}
  {204}},\ \bibinfo {pages} {161} (\bibinfo {year} {2005})}\BibitemShut
  {NoStop}%
\bibitem [{\citenamefont {Elperin}\ \emph {et~al.}(1996)\citenamefont
  {Elperin}, \citenamefont {Kleeorin},\ and\ \citenamefont
  {Rogachevskii}}]{elperin1996turbulent}%
  \BibitemOpen
  \bibfield  {author} {\bibinfo {author} {\bibfnamefont {T.}~\bibnamefont
  {Elperin}}, \bibinfo {author} {\bibfnamefont {N.}~\bibnamefont {Kleeorin}},\
  and\ \bibinfo {author} {\bibfnamefont {I.}~\bibnamefont {Rogachevskii}},\
  }\bibfield  {title} {\bibinfo {title} {Turbulent thermal diffusion of small
  inertial particles},\ }\href@noop {} {\bibfield  {journal} {\bibinfo
  {journal} {Physical review letters}\ }\textbf {\bibinfo {volume} {76}},\
  \bibinfo {pages} {224} (\bibinfo {year} {1996})}\BibitemShut {NoStop}%
\bibitem [{\citenamefont {Grabowski}\ and\ \citenamefont
  {Wang}(2013)}]{grabowski2013growth}%
  \BibitemOpen
  \bibfield  {author} {\bibinfo {author} {\bibfnamefont {W.~W.}\ \bibnamefont
  {Grabowski}}\ and\ \bibinfo {author} {\bibfnamefont {L.-P.}\ \bibnamefont
  {Wang}},\ }\bibfield  {title} {\bibinfo {title} {Growth of cloud droplets in
  a turbulent environment},\ }\href@noop {} {\bibfield  {journal} {\bibinfo
  {journal} {Annual review of fluid mechanics}\ }\textbf {\bibinfo {volume}
  {45}},\ \bibinfo {pages} {293} (\bibinfo {year} {2013})}\BibitemShut
  {NoStop}%
\bibitem [{\citenamefont {Popli}\ \emph {et~al.}(2021)\citenamefont {Popli},
  \citenamefont {Perlekar},\ and\ \citenamefont {Sengupta}}]{popli2020pattern}%
  \BibitemOpen
  \bibfield  {author} {\bibinfo {author} {\bibfnamefont {P.}~\bibnamefont
  {Popli}}, \bibinfo {author} {\bibfnamefont {P.}~\bibnamefont {Perlekar}},\
  and\ \bibinfo {author} {\bibfnamefont {S.}~\bibnamefont {Sengupta}},\
  }\bibfield  {title} {\bibinfo {title} {Pattern stabilization in swarms of
  programmable active matter: A probe for turbulence at large length scales},\
  }\href@noop {} {\bibfield  {journal} {\bibinfo  {journal} {Physical Review
  E}\ }\textbf {\bibinfo {volume} {104}},\ \bibinfo {pages} {L032601} (\bibinfo
  {year} {2021})}\BibitemShut {NoStop}%
\bibitem [{\citenamefont {Crisanti}\ \emph {et~al.}(1990)\citenamefont
  {Crisanti}, \citenamefont {Falcioni}, \citenamefont {Provenzale},\ and\
  \citenamefont {Vulpiani}}]{crisanti1990passive}%
  \BibitemOpen
  \bibfield  {author} {\bibinfo {author} {\bibfnamefont {A.}~\bibnamefont
  {Crisanti}}, \bibinfo {author} {\bibfnamefont {M.}~\bibnamefont {Falcioni}},
  \bibinfo {author} {\bibfnamefont {A.}~\bibnamefont {Provenzale}},\ and\
  \bibinfo {author} {\bibfnamefont {A.}~\bibnamefont {Vulpiani}},\ }\bibfield
  {title} {\bibinfo {title} {Passive advection of particles denser than the
  surrounding fluid},\ }\href@noop {} {\bibfield  {journal} {\bibinfo
  {journal} {Physics Letters A}\ }\textbf {\bibinfo {volume} {150}},\ \bibinfo
  {pages} {79} (\bibinfo {year} {1990})}\BibitemShut {NoStop}%
\bibitem [{\citenamefont {Crisanti}\ \emph {et~al.}(1992)\citenamefont
  {Crisanti}, \citenamefont {Falcioni}, \citenamefont {Provenzale},
  \citenamefont {Tanga},\ and\ \citenamefont
  {Vulpiani}}]{crisanti1992dynamics}%
  \BibitemOpen
  \bibfield  {author} {\bibinfo {author} {\bibfnamefont {A.}~\bibnamefont
  {Crisanti}}, \bibinfo {author} {\bibfnamefont {M.}~\bibnamefont {Falcioni}},
  \bibinfo {author} {\bibfnamefont {A.}~\bibnamefont {Provenzale}}, \bibinfo
  {author} {\bibfnamefont {P.}~\bibnamefont {Tanga}},\ and\ \bibinfo {author}
  {\bibfnamefont {A.}~\bibnamefont {Vulpiani}},\ }\bibfield  {title} {\bibinfo
  {title} {Dynamics of passively advected impurities in simple two-dimensional
  flow models},\ }\href@noop {} {\bibfield  {journal} {\bibinfo  {journal}
  {Physics of Fluids A: Fluid Dynamics}\ }\textbf {\bibinfo {volume} {4}},\
  \bibinfo {pages} {1805} (\bibinfo {year} {1992})}\BibitemShut {NoStop}%
\bibitem [{\citenamefont {Wang}\ \emph {et~al.}(1992)\citenamefont {Wang},
  \citenamefont {Maxey}, \citenamefont {Burton},\ and\ \citenamefont
  {Stock}}]{wang1992chaotic}%
  \BibitemOpen
  \bibfield  {author} {\bibinfo {author} {\bibfnamefont {L.-P.}\ \bibnamefont
  {Wang}}, \bibinfo {author} {\bibfnamefont {M.~R.}\ \bibnamefont {Maxey}},
  \bibinfo {author} {\bibfnamefont {T.~D.}\ \bibnamefont {Burton}},\ and\
  \bibinfo {author} {\bibfnamefont {D.~E.}\ \bibnamefont {Stock}},\ }\bibfield
  {title} {\bibinfo {title} {Chaotic dynamics of particle dispersion in
  fluids},\ }\href@noop {} {\bibfield  {journal} {\bibinfo  {journal} {Physics
  of Fluids A: Fluid Dynamics}\ }\textbf {\bibinfo {volume} {4}},\ \bibinfo
  {pages} {1789} (\bibinfo {year} {1992})}\BibitemShut {NoStop}%
\bibitem [{\citenamefont {Jayaram}\ \emph {et~al.}(2020)\citenamefont
  {Jayaram}, \citenamefont {Jie}, \citenamefont {Zhao},\ and\ \citenamefont
  {Andersson}}]{jayaram2020clustering}%
  \BibitemOpen
  \bibfield  {author} {\bibinfo {author} {\bibfnamefont {R.}~\bibnamefont
  {Jayaram}}, \bibinfo {author} {\bibfnamefont {Y.}~\bibnamefont {Jie}},
  \bibinfo {author} {\bibfnamefont {L.}~\bibnamefont {Zhao}},\ and\ \bibinfo
  {author} {\bibfnamefont {H.~I.}\ \bibnamefont {Andersson}},\ }\bibfield
  {title} {\bibinfo {title} {Clustering of inertial spheres in evolving
  taylor--green vortex flow},\ }\href@noop {} {\bibfield  {journal} {\bibinfo
  {journal} {Physics of Fluids}\ }\textbf {\bibinfo {volume} {32}},\ \bibinfo
  {pages} {043306} (\bibinfo {year} {2020})}\BibitemShut {NoStop}%
\bibitem [{\citenamefont {Baggaley}(2016)}]{baggaley2016stability}%
  \BibitemOpen
  \bibfield  {author} {\bibinfo {author} {\bibfnamefont {A.~W.}\ \bibnamefont
  {Baggaley}},\ }\bibfield  {title} {\bibinfo {title} {Stability of model
  flocks in a vortical flow},\ }\href@noop {} {\bibfield  {journal} {\bibinfo
  {journal} {Physical Review E}\ }\textbf {\bibinfo {volume} {93}},\ \bibinfo
  {pages} {063109} (\bibinfo {year} {2016})}\BibitemShut {NoStop}%
\bibitem [{\citenamefont {Samant}\ \emph {et~al.}(2021)\citenamefont {Samant},
  \citenamefont {Alageshan}, \citenamefont {Sharma},\ and\ \citenamefont
  {Kuley}}]{samant2021dynamic}%
  \BibitemOpen
  \bibfield  {author} {\bibinfo {author} {\bibfnamefont {O.}~\bibnamefont
  {Samant}}, \bibinfo {author} {\bibfnamefont {J.~K.}\ \bibnamefont
  {Alageshan}}, \bibinfo {author} {\bibfnamefont {S.}~\bibnamefont {Sharma}},\
  and\ \bibinfo {author} {\bibfnamefont {A.}~\bibnamefont {Kuley}},\ }\bibfield
   {title} {\bibinfo {title} {Dynamic mode decomposition of inertial particle
  caustics in taylor--green flow},\ }\href@noop {} {\bibfield  {journal}
  {\bibinfo  {journal} {Scientific Reports}\ }\textbf {\bibinfo {volume}
  {11}},\ \bibinfo {pages} {1} (\bibinfo {year} {2021})}\BibitemShut {NoStop}%
\bibitem [{\citenamefont {Maxey}\ and\ \citenamefont
  {Corrsin}(1986)}]{maxey1986gravitational}%
  \BibitemOpen
  \bibfield  {author} {\bibinfo {author} {\bibfnamefont {M.~R.}\ \bibnamefont
  {Maxey}}\ and\ \bibinfo {author} {\bibfnamefont {S.}~\bibnamefont
  {Corrsin}},\ }\bibfield  {title} {\bibinfo {title} {Gravitational settling of
  aerosol particles in randomly oriented cellular flow fields},\ }\href@noop {}
  {\bibfield  {journal} {\bibinfo  {journal} {Journal of Atmospheric Sciences}\
  }\textbf {\bibinfo {volume} {43}},\ \bibinfo {pages} {1112} (\bibinfo {year}
  {1986})}\BibitemShut {NoStop}%
\bibitem [{\citenamefont {Rubin}\ \emph {et~al.}(1995)\citenamefont {Rubin},
  \citenamefont {Jones},\ and\ \citenamefont {Maxey}}]{rubin1995settling}%
  \BibitemOpen
  \bibfield  {author} {\bibinfo {author} {\bibfnamefont {J.}~\bibnamefont
  {Rubin}}, \bibinfo {author} {\bibfnamefont {C.}~\bibnamefont {Jones}},\ and\
  \bibinfo {author} {\bibfnamefont {M.~R.}\ \bibnamefont {Maxey}},\ }\bibfield
  {title} {\bibinfo {title} {Settling and asymptotic motion of aerosol
  particles in a cellular flow field},\ }\href@noop {} {\bibfield  {journal}
  {\bibinfo  {journal} {Journal of Nonlinear Science}\ }\textbf {\bibinfo
  {volume} {5}},\ \bibinfo {pages} {337} (\bibinfo {year} {1995})}\BibitemShut
  {NoStop}%
\bibitem [{\citenamefont {Bergougnoux}\ \emph {et~al.}(2014)\citenamefont
  {Bergougnoux}, \citenamefont {Bouchet}, \citenamefont {Lopez},\ and\
  \citenamefont {Guazzelli}}]{bergougnoux2014motion}%
  \BibitemOpen
  \bibfield  {author} {\bibinfo {author} {\bibfnamefont {L.}~\bibnamefont
  {Bergougnoux}}, \bibinfo {author} {\bibfnamefont {G.}~\bibnamefont
  {Bouchet}}, \bibinfo {author} {\bibfnamefont {D.}~\bibnamefont {Lopez}},\
  and\ \bibinfo {author} {\bibfnamefont {{\'E}.}~\bibnamefont {Guazzelli}},\
  }\bibfield  {title} {\bibinfo {title} {The motion of solid spherical
  particles falling in a cellular flow field at low stokes number},\
  }\href@noop {} {\bibfield  {journal} {\bibinfo  {journal} {Physics of
  Fluids}\ }\textbf {\bibinfo {volume} {26}},\ \bibinfo {pages} {093302}
  (\bibinfo {year} {2014})}\BibitemShut {NoStop}%
\bibitem [{\citenamefont {Govindarajan}(2002)}]{rg2002}%
  \BibitemOpen
  \bibfield  {author} {\bibinfo {author} {\bibfnamefont {R.}~\bibnamefont
  {Govindarajan}},\ }\bibfield  {title} {\bibinfo {title} {Universal behavior
  of entrainment due to coherent structures in turbulent shear flow},\
  }\href@noop {} {\bibfield  {journal} {\bibinfo  {journal} {Physical review
  letters}\ }\textbf {\bibinfo {volume} {88}},\ \bibinfo {pages} {134503}
  (\bibinfo {year} {2002})}\BibitemShut {NoStop}%
\bibitem [{\citenamefont {Maxey}\ and\ \citenamefont
  {Riley}(1983)}]{maxey1983equation}%
  \BibitemOpen
  \bibfield  {author} {\bibinfo {author} {\bibfnamefont {M.~R.}\ \bibnamefont
  {Maxey}}\ and\ \bibinfo {author} {\bibfnamefont {J.~J.}\ \bibnamefont
  {Riley}},\ }\bibfield  {title} {\bibinfo {title} {Equation of motion for a
  small rigid sphere in a nonuniform flow},\ }\href@noop {} {\bibfield
  {journal} {\bibinfo  {journal} {The Physics of Fluids}\ }\textbf {\bibinfo
  {volume} {26}},\ \bibinfo {pages} {883} (\bibinfo {year} {1983})}\BibitemShut
  {NoStop}%
\bibitem [{\citenamefont {Yau}\ and\ \citenamefont
  {Rogers}(1996)}]{yau1996short}%
  \BibitemOpen
  \bibfield  {author} {\bibinfo {author} {\bibfnamefont {M.~K.}\ \bibnamefont
  {Yau}}\ and\ \bibinfo {author} {\bibfnamefont {R.~R.}\ \bibnamefont
  {Rogers}},\ }\href@noop {} {\emph {\bibinfo {title} {A short course in cloud
  physics}}}\ (\bibinfo  {publisher} {Elsevier},\ \bibinfo {year}
  {1996})\BibitemShut {NoStop}%
\bibitem [{\citenamefont {Uhlenbeck}\ and\ \citenamefont
  {Ornstein}(1930)}]{uhlenbeck1930theory}%
  \BibitemOpen
  \bibfield  {author} {\bibinfo {author} {\bibfnamefont {G.~E.}\ \bibnamefont
  {Uhlenbeck}}\ and\ \bibinfo {author} {\bibfnamefont {L.~S.}\ \bibnamefont
  {Ornstein}},\ }\bibfield  {title} {\bibinfo {title} {On the theory of the
  brownian motion},\ }\href@noop {} {\bibfield  {journal} {\bibinfo  {journal}
  {Physical review}\ }\textbf {\bibinfo {volume} {36}},\ \bibinfo {pages} {823}
  (\bibinfo {year} {1930})}\BibitemShut {NoStop}%
\bibitem [{\citenamefont {Childress}(1979)}]{childress1979alpha}%
  \BibitemOpen
  \bibfield  {author} {\bibinfo {author} {\bibfnamefont {S.}~\bibnamefont
  {Childress}},\ }\bibfield  {title} {\bibinfo {title} {Alpha-effect in flux
  ropes and sheets},\ }\href@noop {} {\bibfield  {journal} {\bibinfo  {journal}
  {Physics of the earth and Planetary interiors}\ }\textbf {\bibinfo {volume}
  {20}},\ \bibinfo {pages} {172} (\bibinfo {year} {1979})}\BibitemShut
  {NoStop}%
\bibitem [{\citenamefont {Massot}(2007)}]{massot2007eulerian}%
  \BibitemOpen
  \bibfield  {author} {\bibinfo {author} {\bibfnamefont {M.}~\bibnamefont
  {Massot}},\ }\bibfield  {title} {\bibinfo {title} {Eulerian multi-fluid
  models for polydisperse evaporating sprays},\ }in\ \href@noop {} {\emph
  {\bibinfo {booktitle} {Multiphase reacting flows: modelling and
  simulation}}}\ (\bibinfo  {publisher} {Springer},\ \bibinfo {year} {2007})\
  pp.\ \bibinfo {pages} {79--123}\BibitemShut {NoStop}%
\bibitem [{\citenamefont {De~Chaisemartin}\ \emph {et~al.}(2007)\citenamefont
  {De~Chaisemartin}, \citenamefont {Laurent}, \citenamefont {Massot},\ and\
  \citenamefont {Reveillon}}]{de2007evaluation}%
  \BibitemOpen
  \bibfield  {author} {\bibinfo {author} {\bibfnamefont {S.}~\bibnamefont
  {De~Chaisemartin}}, \bibinfo {author} {\bibfnamefont {F.}~\bibnamefont
  {Laurent}}, \bibinfo {author} {\bibfnamefont {M.}~\bibnamefont {Massot}},\
  and\ \bibinfo {author} {\bibfnamefont {J.}~\bibnamefont {Reveillon}},\
  }\bibfield  {title} {\bibinfo {title} {Evaluation of eulerian multi-fluid
  versus lagrangian methods for ejection of polydisperse evaporating sprays by
  vortices},\ }\href@noop {} {\bibfield  {journal} {\bibinfo  {journal}
  {preprint}\ } (\bibinfo {year} {2007})}\BibitemShut {NoStop}%
\bibitem [{\citenamefont {Hofmann}\ \emph {et~al.}(2018)\citenamefont
  {Hofmann}, \citenamefont {Rieck},\ and\ \citenamefont
  {Sadlo}}]{hofmann2018visualization}%
  \BibitemOpen
  \bibfield  {author} {\bibinfo {author} {\bibfnamefont {L.}~\bibnamefont
  {Hofmann}}, \bibinfo {author} {\bibfnamefont {B.}~\bibnamefont {Rieck}},\
  and\ \bibinfo {author} {\bibfnamefont {F.}~\bibnamefont {Sadlo}},\ }\bibfield
   {title} {\bibinfo {title} {Visualization of 4d vector field topology},\ }in\
  \href@noop {} {\emph {\bibinfo {booktitle} {Computer Graphics Forum}}},\
  Vol.~\bibinfo {volume} {37}\ (\bibinfo {organization} {Wiley Online
  Library},\ \bibinfo {year} {2018})\ pp.\ \bibinfo {pages}
  {301--313}\BibitemShut {NoStop}%
\bibitem [{\citenamefont {Taylor}(1953)}]{taylor1953dispersion}%
  \BibitemOpen
  \bibfield  {author} {\bibinfo {author} {\bibfnamefont {G.~I.}\ \bibnamefont
  {Taylor}},\ }\bibfield  {title} {\bibinfo {title} {Dispersion of soluble
  matter in solvent flowing slowly through a tube},\ }\href@noop {} {\bibfield
  {journal} {\bibinfo  {journal} {Proceedings of the Royal Society of London.
  Series A. Mathematical and Physical Sciences}\ }\textbf {\bibinfo {volume}
  {219}},\ \bibinfo {pages} {186} (\bibinfo {year} {1953})}\BibitemShut
  {NoStop}%
\bibitem [{\citenamefont {Aris}(1956)}]{aris1956dispersion}%
  \BibitemOpen
  \bibfield  {author} {\bibinfo {author} {\bibfnamefont {R.}~\bibnamefont
  {Aris}},\ }\bibfield  {title} {\bibinfo {title} {On the dispersion of a
  solute in a fluid flowing through a tube},\ }\href@noop {} {\bibfield
  {journal} {\bibinfo  {journal} {Proceedings of the Royal Society of London.
  Series A. Mathematical and Physical Sciences}\ }\textbf {\bibinfo {volume}
  {235}},\ \bibinfo {pages} {67} (\bibinfo {year} {1956})}\BibitemShut
  {NoStop}%
\bibitem [{\citenamefont {Young}\ and\ \citenamefont
  {Jones}(1991)}]{young1991shear}%
  \BibitemOpen
  \bibfield  {author} {\bibinfo {author} {\bibfnamefont {W.~R.}\ \bibnamefont
  {Young}}\ and\ \bibinfo {author} {\bibfnamefont {S.}~\bibnamefont {Jones}},\
  }\bibfield  {title} {\bibinfo {title} {Shear dispersion},\ }\href@noop {}
  {\bibfield  {journal} {\bibinfo  {journal} {Physics of Fluids A: Fluid
  Dynamics}\ }\textbf {\bibinfo {volume} {3}},\ \bibinfo {pages} {1087}
  (\bibinfo {year} {1991})}\BibitemShut {NoStop}%
\bibitem [{\citenamefont {San~Miguel}\ and\ \citenamefont
  {Sancho}(1979)}]{san1979brownian}%
  \BibitemOpen
  \bibfield  {author} {\bibinfo {author} {\bibfnamefont {M.}~\bibnamefont
  {San~Miguel}}\ and\ \bibinfo {author} {\bibfnamefont {J.}~\bibnamefont
  {Sancho}},\ }\bibfield  {title} {\bibinfo {title} {Brownian motion in shear
  flow},\ }\href@noop {} {\bibfield  {journal} {\bibinfo  {journal} {Physica A:
  Statistical Mechanics and its Applications}\ }\textbf {\bibinfo {volume}
  {99}},\ \bibinfo {pages} {357} (\bibinfo {year} {1979})}\BibitemShut
  {NoStop}%
\bibitem [{\citenamefont {Rub{\'\i}}\ and\ \citenamefont
  {Bedeaux}(1988)}]{rubi1988brownian}%
  \BibitemOpen
  \bibfield  {author} {\bibinfo {author} {\bibfnamefont {J.~M.}\ \bibnamefont
  {Rub{\'\i}}}\ and\ \bibinfo {author} {\bibfnamefont {D.}~\bibnamefont
  {Bedeaux}},\ }\bibfield  {title} {\bibinfo {title} {Brownian motion in a
  fluid in elongational flow},\ }\href@noop {} {\bibfield  {journal} {\bibinfo
  {journal} {Journal of statistical physics}\ }\textbf {\bibinfo {volume}
  {53}},\ \bibinfo {pages} {125} (\bibinfo {year} {1988})}\BibitemShut
  {NoStop}%
\bibitem [{\citenamefont {Fannjiang}\ and\ \citenamefont
  {Papanicolaou}(1994)}]{fannjiang1994convection}%
  \BibitemOpen
  \bibfield  {author} {\bibinfo {author} {\bibfnamefont {A.}~\bibnamefont
  {Fannjiang}}\ and\ \bibinfo {author} {\bibfnamefont {G.}~\bibnamefont
  {Papanicolaou}},\ }\bibfield  {title} {\bibinfo {title} {Convection enhanced
  diffusion for periodic flows},\ }\href@noop {} {\bibfield  {journal}
  {\bibinfo  {journal} {SIAM Journal on Applied Mathematics}\ }\textbf
  {\bibinfo {volume} {54}},\ \bibinfo {pages} {333} (\bibinfo {year}
  {1994})}\BibitemShut {NoStop}%
\bibitem [{\citenamefont {Subramanian}(2002)}]{subramanian2002inertial}%
  \BibitemOpen
  \bibfield  {author} {\bibinfo {author} {\bibfnamefont {G.}~\bibnamefont
  {Subramanian}},\ }\href@noop {} {\emph {\bibinfo {title} {Inertial effects in
  suspension dynamics}}}\ (\bibinfo  {publisher} {California Institute of
  Technology},\ \bibinfo {year} {2002})\BibitemShut {NoStop}%
\bibitem [{\citenamefont {Bender}\ and\ \citenamefont
  {Orszag}(1978)}]{bender1978sa}%
  \BibitemOpen
  \bibfield  {author} {\bibinfo {author} {\bibfnamefont {C.~M.}\ \bibnamefont
  {Bender}}\ and\ \bibinfo {author} {\bibfnamefont {S.~A.}\ \bibnamefont
  {Orszag}},\ }\bibfield  {title} {\bibinfo {title} {Advanced mathematical
  methods for scientists and engineers},\ }\href@noop {} {\bibfield  {journal}
  {\bibinfo  {journal} {McGraw-Hill, New York}\ }\textbf {\bibinfo {volume}
  {1}},\ \bibinfo {pages} {14} (\bibinfo {year} {1978})}\BibitemShut {NoStop}%
\bibitem [{\citenamefont {Pruppacher}\ and\ \citenamefont
  {Klett}(1978)}]{pruppacher1978microstructure}%
  \BibitemOpen
  \bibfield  {author} {\bibinfo {author} {\bibfnamefont {H.~R.}\ \bibnamefont
  {Pruppacher}}\ and\ \bibinfo {author} {\bibfnamefont {J.~D.}\ \bibnamefont
  {Klett}},\ }\bibfield  {title} {\bibinfo {title} {Microstructure of
  atmospheric clouds and precipitation},\ }in\ \href@noop {} {\emph {\bibinfo
  {booktitle} {Microphysics of Clouds and Precipitation}}}\ (\bibinfo
  {publisher} {Springer},\ \bibinfo {year} {1978})\ pp.\ \bibinfo {pages}
  {9--55}\BibitemShut {NoStop}%
\bibitem [{\citenamefont {Kloeden}\ and\ \citenamefont
  {Platen}(1992)}]{kloeden1992stochastic}%
  \BibitemOpen
  \bibfield  {author} {\bibinfo {author} {\bibfnamefont {P.~E.}\ \bibnamefont
  {Kloeden}}\ and\ \bibinfo {author} {\bibfnamefont {E.}~\bibnamefont
  {Platen}},\ }\bibfield  {title} {\bibinfo {title} {Stochastic differential
  equations},\ }in\ \href@noop {} {\emph {\bibinfo {booktitle} {Numerical
  Solution of Stochastic Differential Equations}}}\ (\bibinfo  {publisher}
  {Springer},\ \bibinfo {year} {1992})\ pp.\ \bibinfo {pages}
  {103--160}\BibitemShut {NoStop}%
\bibitem [{\citenamefont {Vincent}\ and\ \citenamefont
  {Meneguzzi}(1994{\natexlab{b}})}]{vincent1994dynamics}%
  \BibitemOpen
  \bibfield  {author} {\bibinfo {author} {\bibfnamefont {A.}~\bibnamefont
  {Vincent}}\ and\ \bibinfo {author} {\bibfnamefont {M.}~\bibnamefont
  {Meneguzzi}},\ }\bibfield  {title} {\bibinfo {title} {The dynamics of
  vorticity tubes in homogeneous turbulence},\ }\href@noop {} {\bibfield
  {journal} {\bibinfo  {journal} {Journal of Fluid Mechanics}\ }\textbf
  {\bibinfo {volume} {258}},\ \bibinfo {pages} {245} (\bibinfo {year}
  {1994}{\natexlab{b}})}\BibitemShut {NoStop}%
\bibitem [{\citenamefont {Cartwright}\ \emph {et~al.}(2010)\citenamefont
  {Cartwright}, \citenamefont {Feudel}, \citenamefont {K{\'a}rolyi},
  \citenamefont {de~Moura}, \citenamefont {Piro},\ and\ \citenamefont
  {T{\'e}l}}]{cartwright2010dynamics}%
  \BibitemOpen
  \bibfield  {author} {\bibinfo {author} {\bibfnamefont {J.~H.}\ \bibnamefont
  {Cartwright}}, \bibinfo {author} {\bibfnamefont {U.}~\bibnamefont {Feudel}},
  \bibinfo {author} {\bibfnamefont {G.}~\bibnamefont {K{\'a}rolyi}}, \bibinfo
  {author} {\bibfnamefont {A.}~\bibnamefont {de~Moura}}, \bibinfo {author}
  {\bibfnamefont {O.}~\bibnamefont {Piro}},\ and\ \bibinfo {author}
  {\bibfnamefont {T.}~\bibnamefont {T{\'e}l}},\ }\bibfield  {title} {\bibinfo
  {title} {Dynamics of finite-size particles in chaotic fluid flows},\ }in\
  \href@noop {} {\emph {\bibinfo {booktitle} {Nonlinear dynamics and chaos:
  advances and perspectives}}}\ (\bibinfo  {publisher} {Springer},\ \bibinfo
  {year} {2010})\ pp.\ \bibinfo {pages} {51--87}\BibitemShut {NoStop}%
\bibitem [{\citenamefont {Li}\ \emph {et~al.}(2018)\citenamefont {Li},
  \citenamefont {Brandenburg}, \citenamefont {Svensson}, \citenamefont
  {Haugen}, \citenamefont {Mehlig},\ and\ \citenamefont
  {Rogachevskii}}]{li2018effect}%
  \BibitemOpen
  \bibfield  {author} {\bibinfo {author} {\bibfnamefont {X.-Y.}\ \bibnamefont
  {Li}}, \bibinfo {author} {\bibfnamefont {A.}~\bibnamefont {Brandenburg}},
  \bibinfo {author} {\bibfnamefont {G.}~\bibnamefont {Svensson}}, \bibinfo
  {author} {\bibfnamefont {N.~E.}\ \bibnamefont {Haugen}}, \bibinfo {author}
  {\bibfnamefont {B.}~\bibnamefont {Mehlig}},\ and\ \bibinfo {author}
  {\bibfnamefont {I.}~\bibnamefont {Rogachevskii}},\ }\bibfield  {title}
  {\bibinfo {title} {Effect of turbulence on collisional growth of cloud
  droplets},\ }\href@noop {} {\bibfield  {journal} {\bibinfo  {journal}
  {Journal of the Atmospheric Sciences}\ }\textbf {\bibinfo {volume} {75}},\
  \bibinfo {pages} {3469} (\bibinfo {year} {2018})}\BibitemShut {NoStop}%
\bibitem [{\citenamefont {Li}\ \emph {et~al.}(2020)\citenamefont {Li},
  \citenamefont {Brandenburg}, \citenamefont {Svensson}, \citenamefont
  {Haugen}, \citenamefont {Mehlig},\ and\ \citenamefont
  {Rogachevskii}}]{li2020condensational}%
  \BibitemOpen
  \bibfield  {author} {\bibinfo {author} {\bibfnamefont {X.-Y.}\ \bibnamefont
  {Li}}, \bibinfo {author} {\bibfnamefont {A.}~\bibnamefont {Brandenburg}},
  \bibinfo {author} {\bibfnamefont {G.}~\bibnamefont {Svensson}}, \bibinfo
  {author} {\bibfnamefont {N.~E.}\ \bibnamefont {Haugen}}, \bibinfo {author}
  {\bibfnamefont {B.}~\bibnamefont {Mehlig}},\ and\ \bibinfo {author}
  {\bibfnamefont {I.}~\bibnamefont {Rogachevskii}},\ }\bibfield  {title}
  {\bibinfo {title} {Condensational and collisional growth of cloud droplets in
  a turbulent environment},\ }\href@noop {} {\bibfield  {journal} {\bibinfo
  {journal} {Journal of the Atmospheric Sciences}\ }\textbf {\bibinfo {volume}
  {77}},\ \bibinfo {pages} {337} (\bibinfo {year} {2020})}\BibitemShut
  {NoStop}%
\bibitem [{\citenamefont {De~Almeida}(1976)}]{de1976collisional}%
  \BibitemOpen
  \bibfield  {author} {\bibinfo {author} {\bibfnamefont {F.~C.}\ \bibnamefont
  {De~Almeida}},\ }\bibfield  {title} {\bibinfo {title} {The collisional
  problem of cloud droplets moving in a turbulent environinent--part i: A
  method of solution},\ }\href@noop {} {\bibfield  {journal} {\bibinfo
  {journal} {Journal of Atmospheric Sciences}\ }\textbf {\bibinfo {volume}
  {33}},\ \bibinfo {pages} {1571} (\bibinfo {year} {1976})}\BibitemShut
  {NoStop}%
\bibitem [{\citenamefont {De~Almeida}(1975)}]{dealmeida1975effects}%
  \BibitemOpen
  \bibfield  {author} {\bibinfo {author} {\bibfnamefont {F.~C.}\ \bibnamefont
  {De~Almeida}},\ }\bibfield  {title} {\bibinfo {title} {On the effects of
  turbulent fluid motion in the collisional growth of aerosol particles},\
  }\href@noop {} {\bibfield  {journal} {\bibinfo  {journal} {Ph. D. Thesis}\ }
  (\bibinfo {year} {1975})}\BibitemShut {NoStop}%
\bibitem [{\citenamefont {L{\"o}wen}(2020)}]{lowen2020inertial}%
  \BibitemOpen
  \bibfield  {author} {\bibinfo {author} {\bibfnamefont {H.}~\bibnamefont
  {L{\"o}wen}},\ }\bibfield  {title} {\bibinfo {title} {Inertial effects of
  self-propelled particles: From active brownian to active langevin motion},\
  }\href@noop {} {\bibfield  {journal} {\bibinfo  {journal} {The Journal of
  chemical physics}\ }\textbf {\bibinfo {volume} {152}},\ \bibinfo {pages}
  {040901} (\bibinfo {year} {2020})}\BibitemShut {NoStop}%
\bibitem [{\citenamefont {Kobayashi}\ and\ \citenamefont
  {Coimbra}(2005)}]{kobayashi2005stability}%
  \BibitemOpen
  \bibfield  {author} {\bibinfo {author} {\bibfnamefont {M.~H.}\ \bibnamefont
  {Kobayashi}}\ and\ \bibinfo {author} {\bibfnamefont {C.~F.~M.}\ \bibnamefont
  {Coimbra}},\ }\bibfield  {title} {\bibinfo {title} {On the stability of the
  maxey-riley equation in nonuniform linear flows},\ }\href@noop {} {\bibfield
  {journal} {\bibinfo  {journal} {Physics of Fluids}\ }\textbf {\bibinfo
  {volume} {17}},\ \bibinfo {pages} {113301} (\bibinfo {year}
  {2005})}\BibitemShut {NoStop}%
\end{thebibliography}%

\end{document}